\documentclass[ reprint,showpacs,
 superscriptaddress,
 pra,
]{revtex4-1}
\usepackage{mathrsfs,amssymb}
\usepackage{amsmath}
\usepackage{bm}
\usepackage{graphics}
\usepackage{grffile}
\usepackage{empheq}
\usepackage{natbib}
\usepackage{cases}

\renewcommand{\vec}[1]{\bm{\mathrm{#1}}}

\begin{document}

\title{Quantum storage based on the control field angular scanning}
\author{Xiwen Zhang}
\email{xiwen@physics.tamu.edu}
\affiliation{Department of Physics and Astronomy and Institute for Quantum Studies, Texas
A\&M University, College Station, TX 77843-4242, USA}
\author{Alexey Kalachev}
\affiliation{Department of Physics and Astronomy and Institute for Quantum Studies, Texas
A\&M University, College Station, TX 77843-4242, USA}
\affiliation{Zavoisky Physical-Technical Institute of the Russian Academy of Sciences,
Sibirsky Trakt 10/7, Kazan, 420029, Russia}
\author{Olga Kocharovskaya}
\affiliation{Department of Physics and Astronomy and Institute for Quantum Studies, Texas
A\&M University, College Station, TX 77843-4242, USA}
\date{\today }

\begin{abstract}
Continuous change of the propagation direction of a classical control field
in the process of its off-resonant Raman interaction with a weak signal
field in a three-level atomic medium is suggested for quantum storage of a
single-photon wave packet. It is shown that due to phase matching condition
such an angular control allows one to reversibly map the single-photon wave
packet to the Raman spatial coherence grating. Thus, quantum storage and
retrieval can be realized without using inhomogeneous broadening of the
atomic transitions or manipulating the amplitude of the control field. Under
some conditions the proposed scheme proves to be mathematically analogous to
the quantum storage scheme based on controlled reversible inhomogeneous
broadening of the atomic states.
\end{abstract}

\maketitle

{}


\section{Introduction}

Manipulating single-photon states of the electromagnetic field is an
important part of implementation of various quantum information protocols.
In particular, storage and retrieval of single-photon wave packets is of
great importance. Optical quantum memory \cite%
{HSP_2010,TACCKMS_2010,SAAB_2010} is at the heart of linear optical quantum
computing \cite{KMNRDM_2007} and long-distance quantum cryptography with
quantum repeaters \cite{SSRG_2011}. It also may be useful for making
heralded single-photon sources deterministic \cite{PJF_2002}. An efficient
storage and retrieval of single photons has been demonstrated recently in
gases \cite{HSCLB_2011} and rare-earth-ion doped solids \cite{HLLS_2010}
using controlled reversible inhomogeneous broadening (CRIB) of resonant
atomic transitions. Significant progress has also been achieved in
demonstration of quantum storage using atomic frequency comb \cite%
{CRBAG_2010,SBWLAHK_2010,AUALWSSRGK_2010,UARG_2010,SSJSOBGRST_2011,CUBSARG_2011}, electromagnetically induced transparency~ \citep{Chaneliere05, Eisaman05, Novikova07, Choi08} and off-resonance Raman interaction \cite{RNLSLLJW_2010,RMLNLW_2011}.

It is usually assumed that to store and recall optical pulses one needs an
inhomogeneous broadened atomic transition (tailored or controlled) or a
modulated control field amplitude matched an input pulse (see reviews \cite%
{HSP_2010,TACCKMS_2010,SAAB_2010} and recent experiments mentioned above).
In the present work, we develop another approach which requires neither
inhomogeneous broadening nor temporal modulation of the control field
amplitude, but resorts to continuous phase-matching control in an extended
resonant medium. We consider off-resonant Raman interaction of a
single-photon wave packet and a classical control field in a three-level
atomic medium. Under such conditions the phase-matching control can be
achieved by modulating refractive index of the resonant medium %
\citep{Kalachev11} or by modulating the direction of propagation of the
control field, which is studied in the present work. Another possibility is to use a frequency chirp of the control field, which requires synchronous modulation of the atomic transition frequency in order to keep zero two-photon detuning~\citep{ZhangTBP}. In any case, a
continuous change of the wave vector of the control field during the
interaction leads to the mapping of a single photon state to a superposition
of atomic collective excitations (spin waves) with different wave vectors
and vice versa. In comparison with \citep{Kalachev11}, where the
cavity-model of quantum memory was considered, here we discuss a free-space
model. We show that under some conditions the proposed scheme proves to be
mathematically analogous to the longitudinal CRIB-based quantum storage, just as a two-level scheme with refractive index control does ~\citep{Clark12}. The angular control scheme, contrary to CRIB, allows one to use materials which cannot be
controlled by external dc electric or magnetic fields. Generally speaking, the angular control scheme is also much easier for implementation than the refractive index modulation. The problem for the latter is that it needs to be achieved without modulation of atomic levels. It may be done in some specific materials like Tm:LiNbO$_3$~\citep{Clark12}. But in general case such scenario is rather difficult to implement as it was discussed in~\citep{Kalachev11JMO}.
%

The paper is organized as follows. In Sec. \ref{sec:formulism}, the model of
the quantum memory is presented. In Sec. \ref{sec:Single-Mode approximation}%
, we analyze the storage and retrieval of single-photon wave packets with
transverse propagation of the control field. In Sec. \ref{sec:One-Ray
approximation}, a more general situation is considered in the limit of a
large Fresnel number of interaction volume. In Sec. \ref{sec:Implementation
issues}, we consider some implementation issues. Sec. \ref{sec:conclusion}
concludes the paper with final remarks.

\section{The model and basic equations}

\label{sec:formulism}

\begin{figure}[h]
\centering
\resizebox{!}{9.2cm}{\includegraphics{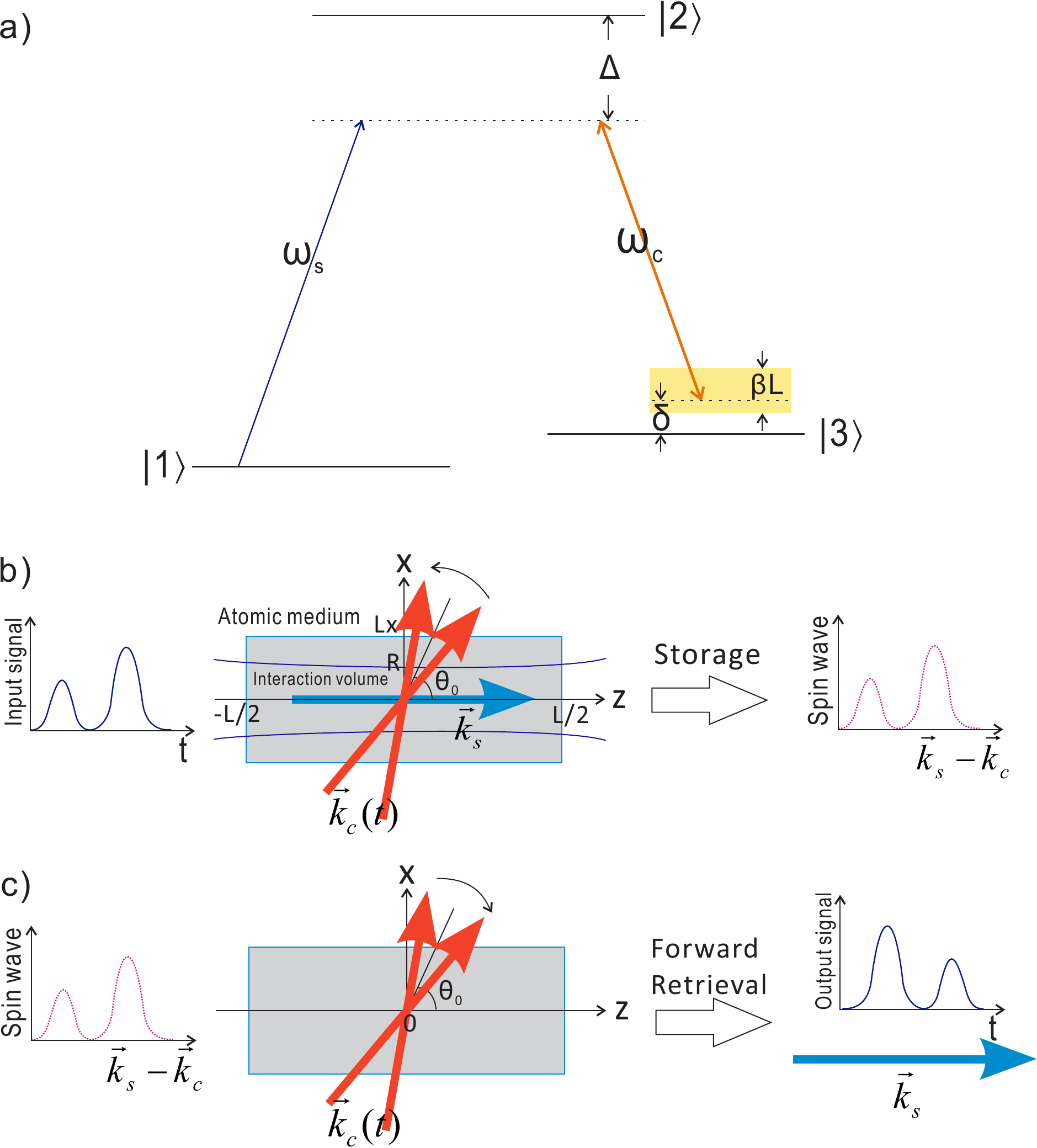}}
\caption{(Color online) (a) Energy diagram of the Raman interaction in a three-level $%
\Lambda $ system; (b) During storage, the temporal profile of the signal
field is mapped into the spin wave distributed over different spin wave
vectors; (c) During forward retrieval, the spin wave profile is mapped back
into the output signal field.}
\label{Figure:Tilt_ Sample}
\end{figure}

We consider a system of identical three-level atoms interacting with a weak
quantum field (single-photon wave packet) to be stored and with a strong
classical control field. The atoms have a $\Lambda $-type level structure,
and the fields are Raman resonant to the lowest (spin) transition, see Fig. %
\ref{Figure:Tilt_ Sample}(a). We assume that the atoms are stationary-like
impurities embedded into a solid-state material or cold atoms in an optical
lattice. The sample is approximated by a parallelepiped with cross section $%
L_{x}\times L_{y}$ and length $L$. The coordinate system is originated at
the center of the medium. The single-photon wave packet is $x$ polarized and
propagates in $z$ direction. It is assumed that the transverse spatial
profile of this quantum field completely goes into the sample cross section.
The control field is supposed to be polarized along axis $y$ and propagated
in any direction in $(x,z)$-plane.
The duration of storage and retrieval processes is equal to $T$. During the
storage, $t\in \left(-T,0\right)$, the propagation direction of the control
field changes so that at different moments, the signal field interacts with
the control field of different directions, thus creates coherence between $%
\left\vert 1\right\rangle $ and $\left\vert 3\right\rangle $ (spin wave)
with different spin wave vectors, see Fig. \ref{Figure:Tilt_ Sample}(b).
During retrieval, $t\in \left(0,T\right)$, the interaction between the spin
wave and the control field reconstructs the signal field, as illustrated in
Fig. \ref{Figure:Tilt_ Sample}(c). The storage time between the end of the
storage and the beginning of the retrieval is subjected a free decay of the
spin wave, which will be neglected through out this paper.

The signal field corresponding to the single-photon wave packet of average
frequency $\omega _{s}$ and wave vector $k_{s}$ is considered in the
paraxial approximation and written as
\begin{equation}
E_{s}(\vec{r},t)=\frac{i}{n}\sqrt{\frac{\hbar \omega _{s}}{2\varepsilon _{0}c%
}}a(\vec{r},t)\,e^{i(k_{s}z-\omega _{s}t)}+\text{H.c.,}
\label{Signal Field Def}
\end{equation}%
where $n$ and $c$ are the refractive index and phase velocity of the quantum
field inside the medium, $a(\vec{r},t)$ is the slowly varying annihilation
operator. Expression (\ref{Signal Field Def}) is written in the
approximation of the equal phase and group speeds of light in the medium
(see, e.g., \citep{Milonni95}) which is made for simplicity.

The classical control field is supposed to be a monochromatic plane wave
subjected to a phase modulation due to rotation of its wave vector on a
small angle. Since most of the atoms remain in the ground state during the
interaction, we neglect the control field absorption so that the field can
be written as
\begin{equation}
E_c(\vec{r},t)=E_0\,e^{i(\bar{\vec{k}}_c\vec{r}-\omega_ct+\phi(\vec{r},t))}+%
\text{c.c.}\,,
\end{equation}
where $E_0$ is a constant amplitude of the plane wave, $\bar{\vec{k}}_c$ is
an average wave vector of $\vec{k}_c(t)$ during rotation, and $\phi(\vec{r}%
,t)$ is a phase shift due to the rotation, which is specified later in this
section.

The collective atomic operators are defined as the mean values of the
single-atom operators
\begin{equation}
\sigma_{mn}(\vec{r},t)=\frac{1}{N}\sum_{j}\lvert m_j\rangle\langle
n_j\rvert\,\delta^{(3)}(\vec{r}-\vec{r}_j)\,,
\end{equation}
where $N$ is the atomic number density, which is supposed to be constant in
space, and $\lvert n_j\rangle$ is the $n$th state ($n = 1,2,3$) of $j$th
atom with the energy $\hbar\omega_n$ ($\omega_1 = 0 < \omega_3 < \omega_2$).
The slowly varying amplitude describing coherence on the Raman transition, $%
s(\vec{r},t)$, is then introduced as
\begin{equation}
\sigma_{13}(\vec{r},t)=s(\vec{r},t)\,e^{i(\vec{k}_s-\bar{\vec{k}}_c)\vec{r}%
-i(\omega_{s}-\omega_{c})t}\,.
\end{equation}

Off-resonant Raman interaction is described by the following equations:
\begin{align}
\left( \frac{\partial }{\partial z}+\frac{1}{c}\frac{\partial }{\partial t}%
\right) a(\vec{r},t) =&\frac{i}{2k_{s}}\Delta _{\perp }a(\vec{r},t)+  \notag
\\
+&g^{\ast }Ns(\vec{r},t)\,e^{i\phi (\vec{r},t)}\,,  \label{A1} \\
\frac{\partial }{\partial t}s(\vec{r},t)=\left( -\gamma +i \delta \right) s&(%
\vec{r},t) -ga(\vec{r},t)\,e^{-i\phi (\vec{r},t)}\,,  \label{A2}
\end{align}%
where $\Delta _{\perp }=\frac{\partial ^{2}}{\partial x^{2}}+\frac{\partial
^{2}}{\partial y^{2}}$, $g=\frac{1}{n}\sqrt{\frac{\omega _{s}}{2\varepsilon
_{0}\hbar c}}\frac{d_{21}\Omega}{\Delta }$ is the coupling constant between
the atoms and the weak quantized field, $\Omega =d_{32}E_{0}/\hbar $ is the
Rabi frequency of the classical control field, $d_{ij}$ is the dipole moment
of the transition between $\left\vert i\right\rangle $ and $\left\vert
j\right\rangle $, $\Delta =\omega _{2}-\omega _{s}$ is one-photon detuning, $%
\gamma $ is the rate of dephasing of the spin coherence, which in the
general case includes both homogeneous and inhomogeneous broadening of the
Raman transition, and $\delta $ is two-photon detuning. 
When writing Eqs.~(\ref{A1}) and (\ref{A2}), it is assumed that 1) the time
of propagation of photons through the system $L/c $ is negligibly short
compared to the evolution time of the slowly time-varying envelopes;
2) all the atoms are initially in the ground state $\lvert 1\rangle $ and
most of them remain in the ground state so that Langevin noise atomic
operators are not included; 3) the frequency shift $|\Omega |^{2}/\Delta $
of the Raman transition induced by the coupling field is taken into account
by redefining field frequency $\omega _{c}$; 4) the refractive index change
of the medium due to the atoms is incorporated into the value of $n$.

It is convenient to expand the slowly varying atomic and field operators on
the transverse mode basis
\begin{align}
a(\vec{r},t)=\sum_{mn}u_{mn}^\ast(x,y)a_{mn}(z,t)\,, \\
s(\vec{r},t)=\sum_{mn}u_{mn}^\ast(x,y)s_{mn}(z,t)\,,
\end{align}
where the transverse mode functions $u_{mn}(x,y)$ satisfy the conditions of
completeness and orthogonality:
\begin{align}
&\sum_{mn} u_{mn}^\ast(x,y)\,u_{mn}(x^{\prime },y^{\prime
})=\delta(x-x^{\prime })\,\delta(y-y^{\prime })\,, \\
&\iint dx\,dy\, u_{mn}^\ast(x,y)\,u_{m^{\prime }n^{\prime
}}(x,y)=\delta_{mm^{\prime }}\,\delta_{nn^{\prime }}\,.
\end{align}
In what follows, we use the set of paraxial plane waves
\begin{equation}
u_{mn}=\frac{1}{\sqrt{L_x L_y}}\,e^{-i\vec{q}_{mn}\boldsymbol{\rho}},
\label{plane wave basis}
\end{equation}
where $\boldsymbol{\rho}=(x,y)$, $\vec{q}_{mn}=(2\pi m/L_x, 2\pi n/L_y)$, $%
m,n\in \mathbb{Z}$, and $L_x, L_y$ are transverse linear sizes of the sample
of a rectangular cross section. This definition corresponds to periodic
boundary conditions in the $x,y$-plane.

In order to write down the equations in the transverse reciprocal space, the
function $\phi(\vec{r},t)$ needs to be specified. Let us suppose that the
wave vector of the control field is rotated around some average direction on
a small angle so that the induced phase shift may be considered as a linear
function of the rotation angle. In such linear regime, we can write
\begin{equation}
\phi(\vec{r},t)=\beta_x(x-x_0)t+\beta_y(y-y_0)t+\beta_z(z-z_0)t+\phi_0(\vec{r%
}),
\end{equation}
where $\beta_i$ is the rate of change of the component $\vec{k}_c$ along $i$%
th axis, and the coordinates $x_0$, $y_0$, $z_0$ correspond to a phase
stationary point where the phase $\phi(t)$ of the control field remains
constant during the rotation. $\phi_0(\vec{r})$ is a time-independent phase
factor which can be incorporated into $s(\vec{r},t)$.

In addition, we extract factors describing the phase due to transverse
momentum by introducing new variables
\begin{align}
A_{mn}(z,t)& =a_{mn}(z,t)w_{mn}(z)\,, \\
S_{mn}(z,t)& =s_{mn}(z,t)w_{mn}(z)\,,
\end{align}%
where $w_{mn}(z)=\exp \left[ i\frac{q_{mn}^{2}}{2k_{s}}(z-z_{w})\right] $,
and $z_{w}$ corresponds to the position of the beam waist. Then the set of
Eqs. (\ref{A1}) and (\ref{A2}) in the reciprocal space take the form
\begin{align}
\left( \frac{\partial }{\partial z}+\frac{1}{c}\frac{\partial }{\partial t}%
\right) & A_{mn}(z,t)=  \notag \\
=g^{\ast }N& \sum_{m^{\prime }n^{\prime }}S_{m^{\prime }n^{\prime
}}(z,t)\,B_{m^{\prime }n^{\prime },mn}^{+}(z,t)\,,  \label{B1} \\
\frac{\partial }{\partial t}S_{mn}(z,t)& =\left( -\gamma +i\delta \right)
S_{mn}(z,t)  \notag \\
-g\sum_{m^{\prime }n^{\prime }}& A_{m^{\prime }n^{\prime
}}(z,t)\,B_{m^{\prime }n^{\prime },mn}^{-}(z,t)\,,  \label{B2}
\end{align}%
where
\begin{multline}
B_{m^{\prime }n^{\prime },mn}^{\pm }(z,t)=\text{sinc}\left[ \frac{(\pm \beta
_{x}t+q_{m^{\prime }n^{\prime },x}-q_{mn,x})L_{x}}{2}\right] \times \\
\text{sinc}\left[ \frac{(\pm \beta _{y}t+q_{m^{\prime }n^{\prime
},y}-q_{mn,y})L_{y}}{2}\right] \times \\
w_{m^{\prime }n^{\prime }}^{\ast }(z)\,w_{mn}(z)\,e^{\pm i(\beta _{z}z-\beta
_{z}z_{0}-\beta _{x}x_{0}-\beta _{y}y_{0})t}\,.
\end{multline}

Without rotating the control field the factors $B^{\pm}_{m^{\prime
}n^{\prime },mn}(z,t)$ become identity matrices, and each transverse mode
can be considered independently in the framework of one-dimensional model.
However, even in the case of rotating control field, the decoupling is also
possible if $\beta_x t$ and $\beta_y t$ remain small with respect to the
interval between the modes $2\pi/L_x$ and $2\pi/L_y$, respectively, while $%
\beta_z t$ being larger than $2\pi/L$. In such a case, different Fourier
transverse modes evolve independently on each other. The quantum storage can
then be considered in single-mode approximation, which is the subject of
Sec. \ref{sec:Single-Mode approximation}. Another simplification is possible
in the limit of a large Fresnel number $F$ of the excitation volume. In this
situation, diffraction spreading may be neglected, and Eqs.~(\ref{A1}) and (%
\ref{A2}) allow one to consider each point $(x,y)$ independently. Such a
geometrical optics approximation is considered in Sec. \ref{sec:One-Ray
approximation}. Regarding the excitation geometry, it should be noted that $%
\beta_x x_0 + \beta_y y_0 + \beta_z z_0$ becomes zero when average control
wave vector $\bar{\vec{k}}_c$ is directed from stationary phase point $%
(x_0,y_0,z_0)$ to the center of the sample. We will see in Sec. \ref%
{sec:One-Ray approximation} that this is the most convenient excitation
geometry, which needs a zero two-photon detuning $\delta $ and requires no
manipulation with $\delta$ during storage and retrieval.

The main figures of merit those should be calculated for the quantum storage
are total efficiency $\eta$ and fidelity $\mathscr{F}$. Since the input and
output faces of the sample are located at the points $z=-L/2$ and $z=L/2$,
respectively, the input and output single-photon wave packets are described
by the operators $A_{mn,\text{in}}(t)=A_{mn}(-L/2,t)$ and $A_{mn,\text{out}%
}(t)=A_{mn}(L/2,t)$. The total efficiency is defined as
\begin{gather}
\eta=\frac{N_\text{out}}{N_\text{in}}\,,
\end{gather}
where
\begin{align}
&N_\text{in}=\sum_{mn}\int_{-\infty}^0 dt \langle A^\dag_{mn,\text{in}%
}(t)A_{mn,\text{in}}(t)\rangle\,, \\
&N_\text{out}=\sum_{mn}\int_{0}^\infty dt \langle A^\dag_{mn,\text{out}%
}(t)A_{mn,\text{out}}(t)\rangle\,,
\end{align}
since we assume that the storage process terminates at the moment $t=0$,
while retrieval process begins at this moment of time.

For the time-reversed output pulse the fidelity is usually defined as
\begin{equation}
\mathscr{F}=\eta \mathscr{F}^{\prime }\,,
\end{equation}%
where
\begin{equation}
\mathscr{F}^{\prime }=\frac{\left\vert \sum_{mn}\int_{0}^{\infty }dt\langle
A_{mn,\text{out}}^{\dag }(t)A_{mn,\text{in}}(\bar{t}-t)\rangle \right\vert
^{2}}{N_{\text{in}}N_{\text{out}}} \label{Fidelity Definition}
\end{equation}
is a measure of pulse preservation independent of the total efficiency, and $%
\bar{t}$ is the delay that maximizes $\mathscr{F}^{\prime }$. In what
follows, it is the latter quantity $\mathscr{F}^{\prime }$ that is used for
characterizing quantum storage together with the efficiency. The definitions
of $\eta $ and $\mathscr{F}^{\prime }$ can also be done in a similar way in
terms of the spatial variables $a_{\text{in}}(\vec{r},t)$ and $a_{\text{out}%
}(\vec{r},t)$. In real space $N_{\text{in}}=\int d^{2}\rho \int_{-\infty
}^{0}dt\,\langle a_{in}^{\dag }(\vec{\rho},t)a_{in}(\vec{\rho},t)\rangle $,
and $N_{\text{out}}=\int_{0}^{\infty }dt\int d^{2}\rho \,\langle a_{\text{out%
}}^{\dag }(\vec{\rho},t)a_{\text{out}}(\vec{\rho},t)\rangle $, while $%
\mathscr{F}^{\prime }=\frac{1}{N_{\text{in}}N_{\text{out}}}\left\vert
\int_{0}^{\infty }dt\int d^{2}\rho \,\langle a_{\text{in}}^{\dagger }(\bar{t}%
-t)a_{\text{out}}(\vec{\rho},t)\rangle \right\vert ^{2}$.

\section{Single-Mode approximation}

\label{sec:Single-Mode approximation}

For simplicity, we assume in what follows that the wave vector of the
control field is rotated in $(x,z)$-plane around an average polar angle $%
\theta_0$, as shown in Fig. \ref{Figure:Wave_vector}, so that $\bar{\vec{k}}%
_{c}=k_{c}\sin \theta_0 \vec{x}+k_{c}\cos \theta_0 \vec{z}$, where $k_{c} =
\frac{\omega_c}{c}$. This also ensures that the control field polarization
is unchanged during the interaction. In this case, we have $\beta _{x}=\beta
\cos \theta_0 $, $\beta _{y}=0$ and $\beta _{z}=-\beta \sin \theta_0 $. The
total angle of rotation $2\Delta \theta$ during the storage or retrieval
process is $2\Delta \theta=\beta T/k_c=\beta T\lambda _{c}/2\pi \ll 1$.

Now we consider the case when the control field propagates perpendicular to
the signal field, i.e. $\theta _{0}=\pi /2$, which corresponds to a
transverse control field. Retrieval is done in a forward way. The transverse
size of our sample is assumed to be
smaller than the sample length. In this case, as pointed out in Sec. \ref%
{sec:formulism}, it is possible to switch between different longitudinal
modes without switching between transverse modes when rotating the control
field. The angle of rotation $\Delta \theta $ should satisfy the following
condition in order to stay within a single transverse mode: $\Delta \theta <%
\sqrt{\frac{2T}{\Delta t}\frac{\lambda _{c}}{L_{x}}}$, where $\Delta t$ is
the duration of the signal pulse. In such a case, Eqs. (\ref{B1}) and (\ref%
{B2}) are decoupled, and $w_{mn}$ gives only diffraction effect for each of
the transverse mode of the signal field. Thus it is sufficient to consider
the evolution of a single mode. Let us define $S_{mn}^{^{\prime
}}=S_{mn}e^{i\beta _{z}\left( z-z_{0}\right) t}$ and go to co-moving frame $%
\tau =t-z/c$. Since $\beta _{x}=0$ and $\beta _{z}=-\beta $, from Eqs.~(\ref%
{B1}) and (\ref{B2}) we have%
\begin{numcases}{}
\frac{\partial }{\partial z}A_{mn}(z,\tau) =g^{\ast }N S_{mn}^{^{\prime }}(z,\tau)\,, \label{One Mode Eqn 1}\\
\frac{\partial }{\partial \tau }S_{mn}^{^{\prime }}(z,\tau) =\left( -\gamma
+i\delta -i\beta (z-z_{0})\right) S_{mn}^{^{\prime }}(z,\tau)-  \notag \\
\qquad \qquad \quad -gA_{mn}(z,\tau)\,. \label{One Mode Eqn 2}
\end{numcases}Then one recognizes that in such a regime, equations
describing the system are the same as those for longitudinal CRIB scheme(
which is also referred to as gradient echo memory scheme)~\citep{Alexander06}.
In longitudinal CRIB scheme, a space dependent absorption line along the
medium is created. While the signal field propagates through the medium,
different frequency components get absorbed by different absorption lines at
different longitudinal positions, resulting in a space dependent coherence
in the sample. During retrieval, such coherence is mapped back onto the
output signal. We can understand our scheme in exactly the same way: Since
the rotation of the control field yields a factor $e^{i\phi (\vec{r}%
,t)}=e^{i\beta _{z}t(z-z_{0})}$, thus on the one hand, there is a time
dependent wave vector $\beta t$ which is responsible for the writing of the
spin waves with different wave vectors and the recording of the temporal
profile of the signal field. On the other hand, this term can be viewed as a
space dependent absorption line at frequency $\beta z$, which absorbs
different frequency components of the signal at different positions of the
medium. The medium opens an absorption window of width $\beta L$ (Fig. \ref%
{Figure:Tilt_ Sample}(a)) along longitudinal direction. The absorption of
the central frequency of the signal field happens at $z_{p}=z_{0}+\delta
/\beta $. It is necessary for $z_{p}$ to be inside the medium, and better at
the middle of the sample. This means that if $z_{0}=0$, i.e. the
longitudinal position of the phase stationary point of the control field
corresponds to the center of the sample, then the two-photon detuning $%
\delta $ should be equal to zero. Otherwise, one can take advantage of the
two-photon detuning to shift the phase stationary point longitudinally to
the medium center. Although the equations of motion can be reduced to the same as those of CRIB, the underlying physics of our scheme is quite different. Instead of controlling inhomogeneous broadening, we realize quantum storage by continuously creating spin wave vectors of different values to record the temporal information of the incoming single-photon wave packet. In other words, we achieve the same storage effect as in CRIB via a specific phase control of the control field which results in the same effect as the spatially depended frequency control of the atomic levels.

In order to have good storage efficiency, the absorption window width $\beta
L$ should cover the input pulse spectrum width: $\beta L > 2 \pi / \Delta t$%
. So
\begin{equation}
2 \Delta \theta > \frac{T}{\Delta t}\frac{\lambda _{c}}{L}\, .
\label{Window condition 2}
\end{equation}%
This condition is the same as that resulted from switching between different
longitudinal modes during the storage of the signal pulse. On the other
hand, it is known that the ``optical density'' for each spectral component
of the signal field in Fourier space for longitudinal CRIB~\citep{Longdell08}
is $2\pi \left\vert g\right\vert ^{2}N/\beta $. This quantity needs to be
larger than unity, so%
\begin{equation}
2 \Delta \theta < T\lambda _{c}\left\vert g\right\vert ^{2}N \, .
\label{Absorption condition}
\end{equation}%
Thus one needs to keep a balance between the absorption window width and the
``optical density''. The bigger $\beta $ is, the wider the absorption window
opens while the smaller the ``optical density'' becomes, and vice versa.
Besides, we choose the parameters to avoid significant storage phase
factor~\citep{Longdell08} throughout the paper. The retrieval of the
signal is done by switching the sign of $\beta $ (and the sign of $\delta $,
if a non-vanishing $\delta $ is used to shift the absorption center),
corresponding to a reversal scan of the control field. The retrieval field experiences a phase modulation~\citep{Moiseev08}, which can strongly decrease the fidelity defined as (\ref{Fidelity Definition}), and has to be taken into account beyond conditions (\ref{Window condition 2}) and (\ref{Absorption condition}). The performance of
the storage and retrieval is then equivalent to that of longitudinal CRIB,
which has been well studied and understood in many papers~ %
\citep{Hetet08,Hetet08EXP,Hetet08EXP3level,Moiseev08}. However, it is worth
noting that a retrieval signal can also be generated without switching the
scanning direction of the control field (thus the sign of $\beta$), as
opposed to longitudinal CRIB scheme. Yet in this paper, we always perform a
reversal scan during retrieval.

We note here that Eqs. (\ref{One Mode Eqn 1}) and (\ref{One Mode Eqn 2})
with $z_0 = -L/2$ are also the equations describing quantum memory scheme
via refractive index control~\citep{Kalachev11} in free space, for which the
above discussions are applicable. In this case, a non-vanishing two-photon
detuning $\delta =\beta L/2$ is required to shift the position of the
absorption line of the central frequency of the signal field to the center
of the medium.

\section{Geometrical Optics Approximation}

\label{sec:One-Ray approximation}

In the previous section, we consider a special case when the propagation
directions of the signal and the control fields are perpendicular to each
other. This allows us to decouple equations in the reciprocal space and make
use of single-mode approximation. However, the transverse excitation
requires more power because of the large control beam cross section. In this
section we consider control field propagating at an arbitrary angle with
respect to the $z$ axis. But we make use another approximation when the
Fresnel number $F$ of interaction volume is much bigger than unity. Then the
second order transverse derivative in Eqs.~(\ref{A1}) can be dropped so that
each point in the transverse plane $(x,y)$ can be considered independently.
The present case corresponds to geometrical optics when the signal field is
described as a series of rays propagating along the axis $z$. The transverse
profile of the signal field is assumed to be smooth enough and treated as
featurelessness within each of the Fresnel zones.

\begin{figure}[h]
\centering
\resizebox{!}{8.9cm}{\includegraphics{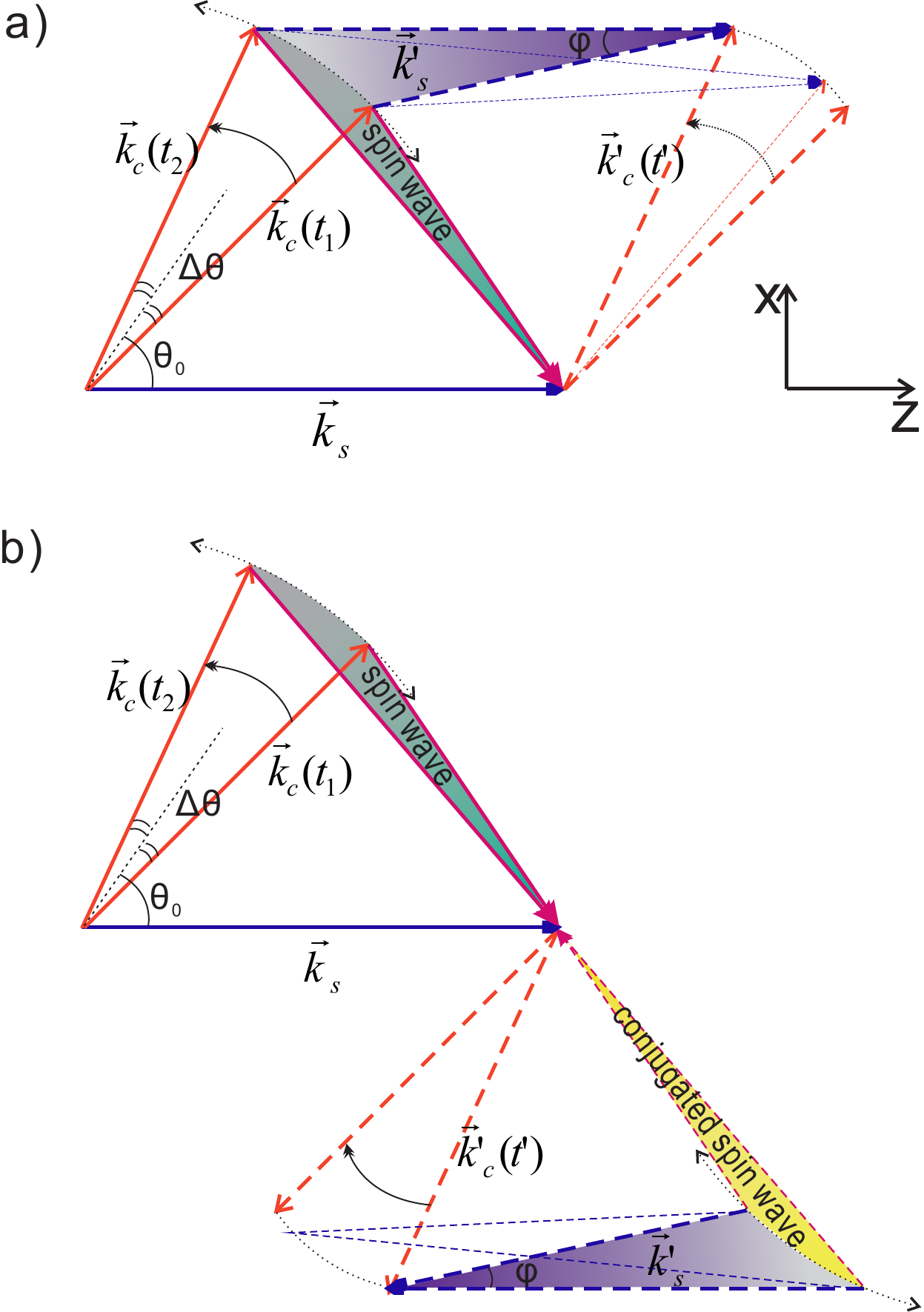}}
\caption{(Color online) Diagram of the signal wave vector, control wave vector and spin
wave vector for (a) forward retrieval and (b) backward retrieval. A writing
control wave vector changes direction from $\vec{k}_{c}(t_1)$ to $\vec{k}%
_{c}(t_0)$. For $t_1=-T$, $t_2=0$, the control wave vector sweeps an angle $%
2\Delta \protect\theta$ around an average polar angle $\protect\theta_0$.
When a reading control field is applied at time $t^{\prime}$, the retrieved
signal field spreads over an angle $\protect\varphi$. In backward retrieval,
the spin waves need to be flipped to their conjugated components in the time
interval between the end of the storage and beginning of the retrieval.}
\label{Figure:Wave_vector}
\end{figure}

In the reciprocal space, the second order transverse derivative term $\frac{i%
}{2k_{s}}\Delta _{\perp }{a}\left( \vec{r},t\right) $ gives $%
-iq_{mn}^{2}/\left( 2k_{s}\right) a_{mn}\left( z,t\right) $. This term
describes phase shift due to transverse momentum, which can be ignored when $%
q_{mn}^2/k_s\ll \pi$. The latter corresponds to a very directional
propagation of the signal field when Fresnel number of interaction volume is
large. However, when the control field is rotated and different spin waves
are created in the medium during the Raman interaction, the signal beam can
be subject to spreading in transverse direction due to the scattering of the
control field on the spin waves with different wave vectors. Therefore, in
order to neglect the transverse derivative in the equations, we need not
only large Fresnel number of the interaction volume, but also sufficiently
small angle of control field rotation $\Delta \theta$. The upper limit of $%
\Delta \theta$ can be estimated in the following way. Consider the
interaction between signal field of wave vector $\vec{k}_{s} $ and control
field of wave vector $\vec{k}_{c}(t)$ (Fig. \ref{Figure:Wave_vector}). While
$\vec{k}_{c}$ changes from $\vec{k}_{c}(t_{1}) $ (at an angle $\theta_0
-\Delta \theta$) to $\vec{k}_{c}(t_{2}) $ (at an angle $\theta_0 +\Delta
\theta$), it continuously creates a set of spin wave vectors $\vec{k}_{s}-%
\vec{k}_{c}(t)$ depicted in Fig. \ref{Figure:Wave_vector}. As a result the
signal wave vector can spread over an angle $\varphi $. For $k_{s}\approx
k_{c}$, $\Delta \theta\ll 1$, one can estimate $\varphi <2\Delta \theta$. We
need $\left( \Delta q_{x}^{2}+\Delta q_{y}^{2}\right) L/\left( 2k_{s}\right)
\ll \pi $. Since $\Delta q_{x}\sim k_{s}\sin \varphi \leqslant 2\Delta
\theta k_{s}$, $\Delta q_{y}=0$, we arrive eventually at the following
limit: $\Delta \theta\ll \Delta\theta _{\max }=\frac{1}{2} \sqrt{\frac{%
\lambda _{s}}{L}} \approx \frac{1}{2}\sqrt{F}\theta_d$, where $\theta_d$ is
the diffraction angle. It should be noted that this inequality can be
considered as just a sufficient condition for neglecting second derivatives
and geometrical optics approximation, but not necessary for the considered
quantum storage protocol.

Now, by defining a new variable $s^{{\prime }}(\vec{r},t)=s(\vec{r},t)\cdot$
$\cdot e^{i \left[\beta _{x}\left( x-x_{0}\right) t + \beta _{z}\left(
z-z_{0}\right) t \right]}$,\ we have the following equations in the
co-moving frame:%
\begin{numcases}{}
\frac{\partial }{\partial z}a(\vec{r},\tau) =g^{\ast }Ns^{\prime }(\vec{r}%
,\tau)\,,  \label{One-ray A 1} \\
\frac{\partial }{\partial \tau}s^{\prime }(\vec{r},\tau) =\{-\gamma +i\delta
+i[\beta \cos \theta_0 \left( x-x_{0}\right) -  \notag \\
\qquad \quad -\beta \sin \theta_0 \left( z-z_{0}\right) ]\}s^{\prime }(\vec{r},\tau)-ga(\vec{%
r},\tau)\,.  \label{One-ray S 1}
\end{numcases}

Taking the same argument as in Sec. \ref{sec:Single-Mode approximation}, we
arrive at the following condition for efficient quantum storage:
\begin{equation}
\frac{T}{\Delta t}\frac{\lambda _{c}}{L} < 2 \Delta \theta \sin \theta_0
< T\lambda _{c}\left\vert g\right\vert ^{2}N.
\label{one-ray efficiency and fidelity cond}
\end{equation}%
Beyond this, we need to confine the value of the average polar angle $%
\theta_0 $. The central absorption line appears at the position $z_{p}=\frac{%
\delta }{\beta \sin \theta_0 }+z_{0}+\cot \theta_0 \left( x-x_{0}\right) $. $%
z_{p}$ should be inside the medium, so%
\begin{equation}
-L/2<z_{p}<L/2.  \label{zp condition}
\end{equation}%
Again, one can engineer the control field to make the location of the phase
stationary point $\left( x_{0},z_{0}\right) =\left( 0,0\right) $ and set
two-photon detuning $\delta =0$. However, this condition can be relaxed in a
way that the frequency shift introduced by $z_{0}$ is canceled by the shift
introduced by\ $x_{0}$ and $\delta $, namely, $z_{0}-x_{0}\cot \theta_0
+\delta /\left( \beta \sin \theta_0 \right) =0$. So the phase stationary
point does not need to be exactly at the center of the sample, but may be situated anywhere on the line%
\begin{equation}
x_{0}=z_{0}\tan \theta_0 +\delta /\left( \beta \cos \theta_0 \right) \,.
\label{position of phase stationary point}
\end{equation}%
When $\delta =0$, this is just the bisector of the rotation. If there exists
any difficulty to put $\left( x_{0},z_{0}\right) $ on the bisector, a
non-vanishing $\delta $ can be used to shift the position of the absorption
lines. In the following we will assume this condition is fulfilled. Then (%
\ref{zp condition}) gives $-L/2<x\cot \theta_0 <L/2$. Replace $x$ by the
radius of the excitation volum $R$, we have $\left\vert \cot \theta_0
\right\vert <\frac{L}{2R}$. Define geometry angle $\theta_{g}=\arctan \frac{%
2R}{L}$, then the condition on the average polar angle $\theta_0 $ can be
written as
\begin{equation}
\theta_0 >\theta _{g}=\arctan \frac{2R}{L} \,.  \label{theta condition}
\end{equation}%
If (\ref{position of phase stationary point}) is not fulfilled, the angle $%
\theta_0 $ needs to be even bigger. Since $\theta _{g}\approx F\theta _{d}$
and $\Delta \theta _{\max }\approx \frac{1}{2}\sqrt{F}\theta _{d}$, for
Fresnel number $F\gg 1$ condition (\ref{theta condition}) automatically
ensures that $\theta_0 \gg \Delta \theta$, which is the condition we have
been using through out this paper.

\subsection{Forward retrieval}

\begin{figure}[h]
\centering
\resizebox{!}{7.0cm}{
\includegraphics{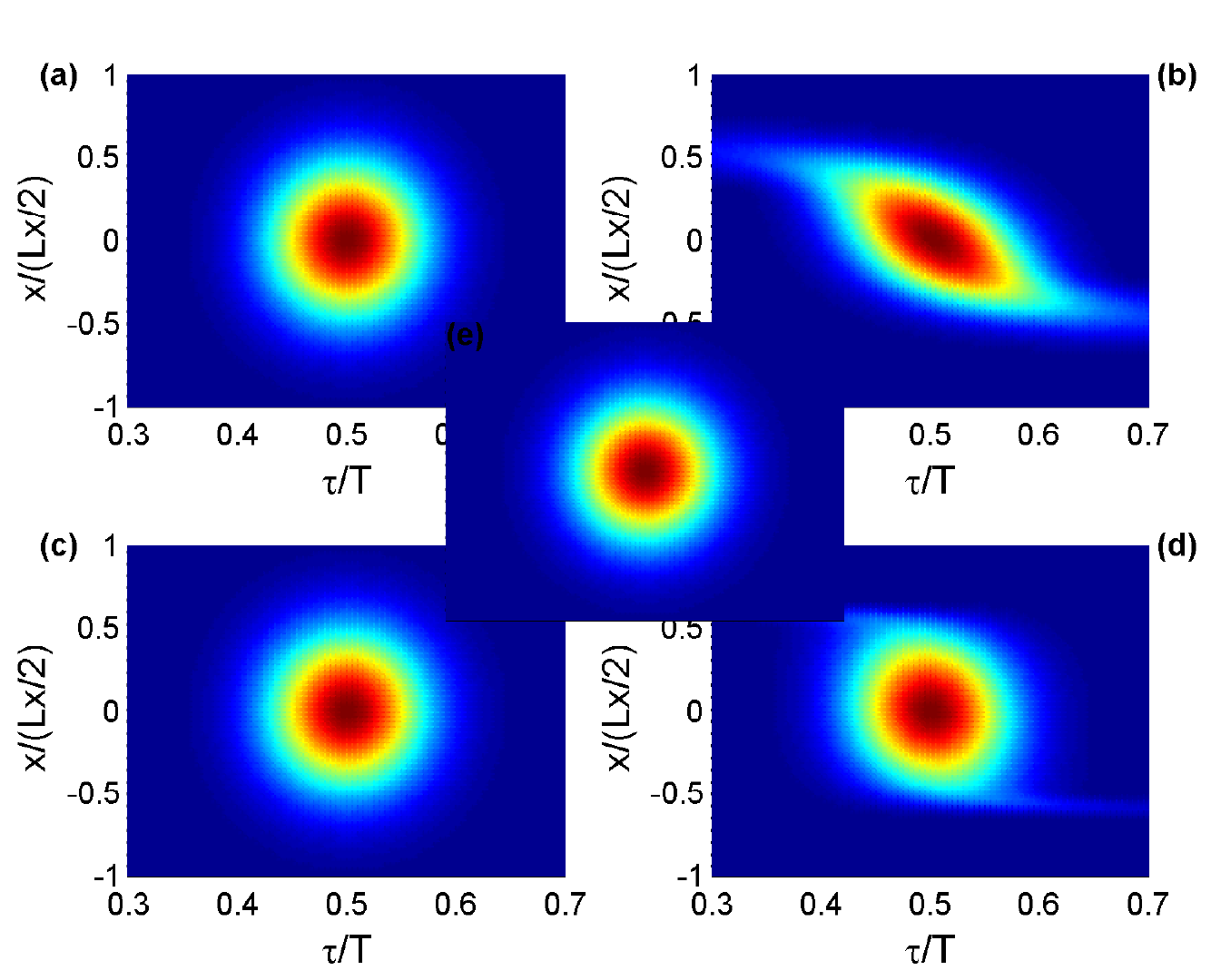}}
\caption{(Color online) Forward retrieval fields for $\protect\theta_0=\protect\pi/2$ in
(a) $\&$ (c) and $\protect\theta_0=\protect\pi/9$ in (b) $\&$ (d). (e) shows
the signal field to be stored. The vertical axis $x/(L_x/2)$ is the transverse dimension of the sample normalized by the half transverse size $L_x/2$; the horizontal axis $\tau /T$ is the local time normalized by the duration of the storage(retrieval) process $T$. The color indicates the signal field amplitude. (a, b) Retrieval fields for constant ${\Delta
\protect\theta}$. The field is distorted for small $\protect\theta_0$
because (1): the absolute value of ${\protect\beta_z = - \protect\beta \sin\protect\theta_0}$
becomes smaller such that it approaches the lower limit of (\protect\ref%
{one-ray efficiency and fidelity cond}); (2): ${\protect\beta_x = \protect%
\beta \cos\protect\theta_0}$ becomes large enough to lead a spacial-temporal
distortion; (c, d) Retrieval fields for constant ${\Delta \protect\theta \sin%
\protect\theta_0}$. There is transverse distortion of the retrieval field
due to the non-vanishing ${\protect\beta_x}$. The figures are generated
under the following parameters: ${\protect\lambda_s\approx\protect\lambda%
_c=1550}$~nm( fiber-optic communication band), ${|g|^2 N = 8.3\times
10^{10}}$~/(s$\cdot$~m), ${\Delta t/T = 1/20}$, ${T=1000}$~ns, ${2R/L_x=1/6}$%
, where $R$ is considered to be the transverse spatial half width of the
input signal, ${L_x=0.6}$~cm, ${L=1}$~cm, ${\Delta \protect\theta = 8\times
10^{-3} }$~rad in (a) $\&$ (b), ${{= 8\times 10^{-3}/\sin\protect\theta_0 }}$~rad
in (c) $\&$ (d). }
\label{Figure:Forward_Field_Distoration}
\end{figure}

Under the condition (\ref{position of phase stationary point}), the Eqs. (%
\ref{One-ray A 1}) and (\ref{One-ray S 1})\ become 
\begin{numcases}{}
\frac{\partial }{\partial z}a(\vec{r},\tau) =g^{\ast }Ns^{\prime }(\vec{r}%
,\tau)\,, \\
\frac{\partial }{\partial \tau}s^{\prime }(\vec{r},\tau) =[-\gamma +i\beta (x\cos
\theta_0 -  \notag \\
\qquad \quad \quad \  -z\sin \theta_0 )]s^{\prime }(\vec{r},\tau)-ga(\vec{r},\tau)\,.
\end{numcases}The term $\beta x\cos \theta_0 s^{\prime }(\vec{r},t)$ has no
positive contribution to the storage and retrieval of the signal field. It
appears as a side effect due to the change of $\vec{k}_{c}$ on transverse
direction. For forward retrieval, this term could introduce some transverse
profile to the retrieved signal. This transverse profile reveals itself as a
transversal space dependent time shift (ahead or behind) of the forward
retrieved signal, as seen in Fig. \ref{Figure:Forward_Field_Distoration}
(d).
The reason of the time delay is as follows: the signal field in the medium
excites spin wave and evolves in the form of polariton, which is the
combination of the signal field and the excited spin wave~\citep{Hetet08}.
The polariton, after storage, freezes inside the medium at the position of
central absorption line $z_{p}=x\cot \theta_0 $. It is now clear that the
condition (\ref{zp condition}) is to place the polariton just inside the
medium. Otherwise if no polariton is created, the signal field can not be
stored. During forward retrieval, the control field picks up the polariton
at $z_{p}$ and generates retrieval signal at $z=L/2$. So the group velocity
of the signal field reduces from $c$ to $0$ while $z=-L/2\rightarrow z_{p}$,
and recovers from $0$ to $c$ while $z=z_{p}\rightarrow L/2$. As a result it
is obvious that if $z_{p}$ is not exactly equal to $0$, there must be a time
shift going along with the forward retrieved signal. This time shift is
avoided in the case of $\theta_0 =\pi /2$ because there $z_{p}$ is
independent on $x$ (transversely homogeneous). Even if $z_{p}$ is not
exactly equal to $0$, there would be only a time shift as a whole for the
recalled signal. However, for $\theta_0 \neq \pi /2$, at $x \neq 0$ this
time shift is unavoidable for forward retrieval, and moreover, it must
introduce a transverse distortion of the signal. For $z_{p}$ not deviating
much from $0$ and $4c\left\vert g \right\vert ^2 N / (\beta^2 L^2
\sin^2\theta_0) \gg 1$, the group velocity of the signal field near the ends
of the medium is~\citep{Moiseev08} $v_{g}(z)|_{\left\vert z \pm
L/2\right\vert \ll L/2 }\sim \beta ^{2}\left( z-z_{p}\right) ^{2}/\left(
\left\vert g\right\vert ^{2}N\right) $. Since the delay is due to the
propagation between $z=L/2 + 2z_{p}$ and $z=L/2$, from $dz=v_{g}(z)dt$, the
delay time can be estimated as
\begin{equation}
t_{d}\sim \frac{\left\vert g\right\vert ^{2}N}{\beta ^{2} \sin^2\theta_0}%
\left( \frac{1}{L/2+z_{p}}-\frac{1}{L/2-z_{p}}\right) \,.  \label{td}
\end{equation}%
Taking $z_{p}=\pm R\cot \theta_0 $, one estimates the temporal broadening of
the forward retrieved signal with respect to the input signal as
\begin{equation}
\Delta t_{br}\sim \frac{\left\vert g\right\vert ^{2}N}{\beta
^{2}\sin^2\theta_0}\frac{4R\cot \theta_0 }{\left( L/2\right) ^{2}-R^{2}\cot
^{2}\theta_0 }\,,  \label{tbroaden}
\end{equation}%
which is finite and positive considering the condition (\ref{theta condition}%
).
\begin{figure}[h]
\centering
\resizebox{!}{6.9cm}{\includegraphics{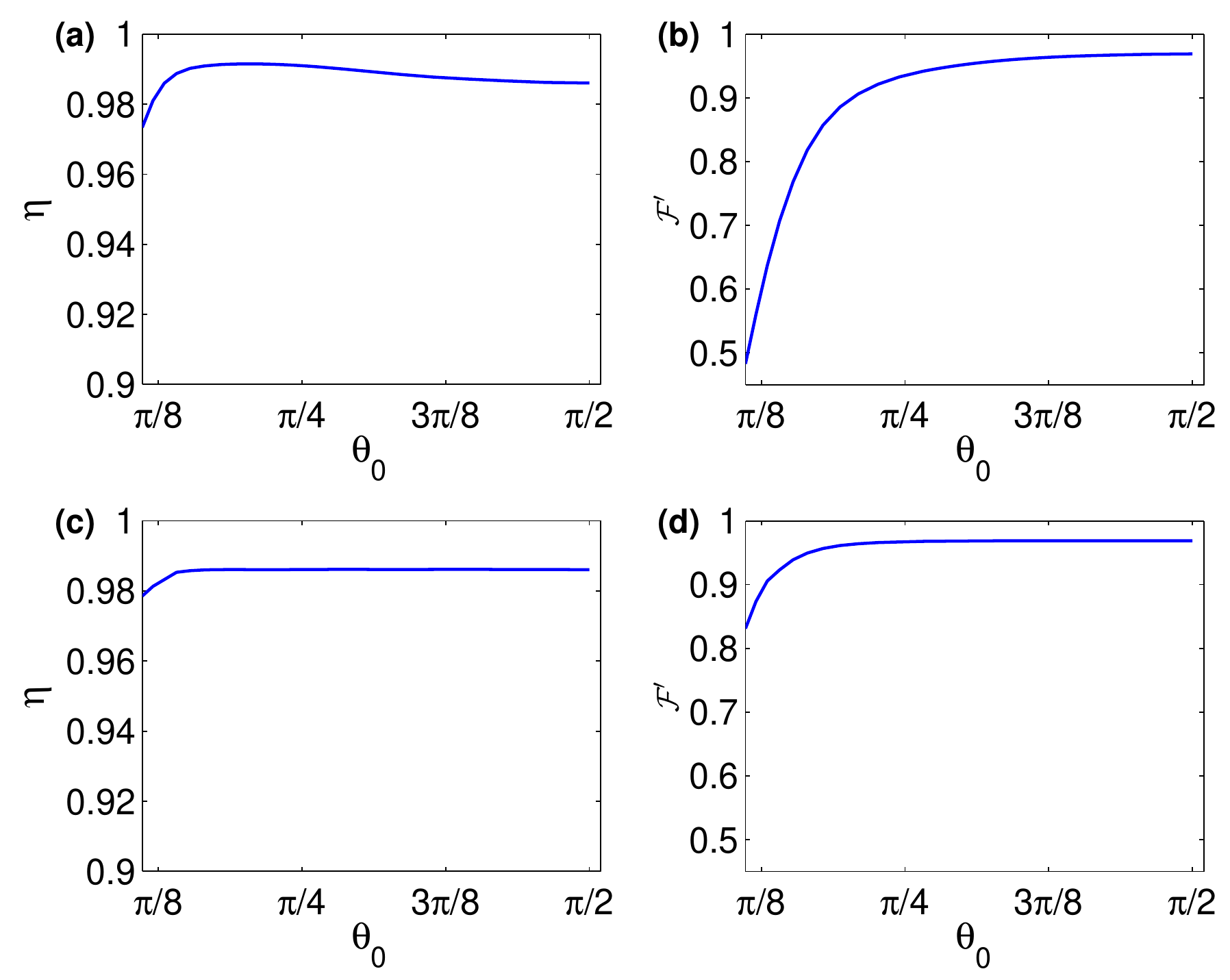}}
\caption{(Color online) Forward retrieval Efficiency $\protect\eta$ and Fidelity $
\mathscr{F}^{\prime}$ v.s. the average polar angle $\protect\theta_0$. (a,
b) Efficiency(a) and Fidelity(b) for forward retrieval with ${\Delta \protect%
\theta}$ kept as constant; (c, d) Efficiency(c) and Fidelity(d) for forward
retrieval with ${\Delta \protect\theta \sin \protect\theta_0}$ kept as
constant. The parameters are the same as described in Fig. \protect\ref%
{Figure:Forward_Field_Distoration}. }
\label{Figure:Eff_Fid}
\end{figure}

Fig. \ref{Figure:Eff_Fid} shows how the efficiency and fidelity depend on $%
\theta_0 $. Although a smaller $\theta_0 $ requires less control field
power, it gives less fidelity due to the transversal distortion for a given $%
\Delta \theta$. Alteratively, it is necessary to increase $\Delta \theta$
when $\theta_0 $ approaches $\theta_g$ from above, in order to maintain high
fidelity of quantum storage, as shown in Fig. \ref{Figure:Eff_Fid} (c, d).

\subsection{Backward retrieval}

In the case of forward retrieval, the above transverse distortion decreases
the fidelity of quantum storage when the average polar angle $\theta_0$
becomes small. This can be avoided in a backward retrieval, wherein the time
shift at each transverse point is exactly compensated during backward
propagation. As a result, the fidelity remains high for all average angles $%
\theta_0$ of interest. However, due to the phase matching condition, it is
not enough to merely switch the propagation direction of the control field
to the opposite. In general, it requires phase conjugation of the spin waves.
Under considered geometrical optics approximation, it is sufficient to flip
the direction of the spin wave vector along direction
$\vec{K} = \vec{k}_s-\vec{k}_c(t=0)$ before the retrieval. This can be done
by applying two non-colinearly propagating $\pi $ pulses one followed by
another~\citep{Moiseev01,Nilsson05}, both on resonance with level $%
\left\vert 3\right\rangle $ and another energy level, say, $\left\vert
4\right\rangle $. The transition frequency $\omega_{34}$ needs to be bigger
than $c \lvert \vec{K}\rvert$. The directions of the two $\pi$ pulses are
arranged in such a way that the wave vectors difference between them is
along $-\vec{K}$. Another possibility is to generate spin wave vectors
perpendicular to the signal wave vector~\citep{Surmacz08}. In this case, $%
k_{c}$ and $k_{s}$ needs to differ from each other significantly.

In the case of backward retrieval, the input signal enters at $z=-L/2$ and
creates polariton at $z=z_{p}$, while the retrieved signal starts from $%
z=z_{p}$ and emits at $z=-L/2$. As a result, there is no delay of the output
field for any value of $z_{p} \in (-L/2,L/2)$.

Generally speaking, in either forward or backward retrieval, the scanning of the control field can be implemented in different ways. The control of the scanning order and scanning rates offers the possibility of the manipulation of the retrieval pulses, for instance, recalling different temporal parts of the input pulses in different orders. Such a manipulation in the angular-time domain may be thought of as counterpart of that in the frequency-spatial domain using CRIB~\citep{Buchler11}.

\section{Implementation issues}

\label{sec:Implementation issues}

Let us discuss possible experimental conditions under which the storage and
retrieval of weak optical pulses via angular scanning of the control field
is possible.

In Fig. \ref{Figure:Forward_Field_Distoration} and \ref{Figure:Eff_Fid} we
demonstrate the performance of our quantum memory scheme with signal field
frequency around fiber-optic communication band. Yet experimentally it is
preferable to store and recall the signal field with small total angle of
rotation $2\Delta \theta$ and rotation rate $2\Delta \theta/T$. If we take ${\lambda_s \approx \lambda_c=700}$%
~nm, according to the numerical simulations, a single
Gaussian pulse of duration $\Delta t=T/20$ with $T=500$~ns can be stored and recalled with
high efficiency and fidelity using transverse control field if the angle of
rotation is of the order of $10\lambda/L$, which gives $\Delta \theta / T \sim 10^3\text{~rad/s}$ for ${L=1}$~cm. Such rate of beam deviation can be achieved by
commercial equipment. Regarding promising storage materials, two points must
be kept in mind in the context of the present scheme: 1) off-resonant Raman
interaction needs systems with relatively strong optical transitions,
especially in the case of transverse control field; 2) exploiting coherent
spatial grating needs atoms to be stationary in space. From this point of
view, a system of cold atoms trapped in an optical lattice seems to be one of the
promising candidates \citep{STTPDKB_2009,DZKK_2010}. The influence of
regular atomic structure on the efficiency of quantum memory in such a
system was analyzed recently in \citep{NDMRLLWJ_2010}. We note only that the
maximum length of the wave vector of a spin wave created via non-collinear
Raman interaction is limited by the minimum interatomic distance, which
means that a blue detuned optical lattice should be used for the case of
transverse control field. Another promising material is an ensemble of
defect centers in diamond such as nitrogen-vacancy centers \citep{JW_2006,
BNTMKMIABTJHJW_2009}. The existence of lambda type optical transitions in
such a system was demonstrated via electromagnetically induced transparency %
\citep{HTSM_2001} and coherent population trapping \citep{STNWFBROGPJH_2006}%
. The effect of inhomogeneous broadening of the Raman transition may be
removed through the use of a spin-echo pulse sequence, and long dephasing
times are achieved by dynamic decoupling of spin qubits from their local
environment \citep{LWRDH_2010,RHC_2010,NDHSFHJW_2011}. Moreover, transfer of
quantum states between the electron spins and nuclear spins is possible %
\citep{GCJTMJZHL_2007}, which in combination with dissipative decoupling
schemes allows long-lived quantum storage \citep{MKLJYBPHCMTCL_2012}.

\section{Conclusion}

\label{sec:conclusion}

We propose a new method to store and retrieve weak pulses such as
single-photon wave packets based on off-resonant Raman interaction. By
changing the propagation direction of the strong classical control field,
the temporal profile of the signal field is mapped into the spatial grating
of Raman coherence. If the wave vector of the control field is perpendicular
to that of the signal field, the quantum storage can be described in
single-mode approximation so that the proposed scheme proves to be
equivalent to longitudinal CRIB scheme. When the control field
approaches longitudinal one, in the case of forward retrieval there maybe
additional spatial-temporal distortion of the output field with respect to
the input signal due to the change of the control wave vector on transverse direction. Such distortion can be
avoided in backward retrieval. The proposed scheme has the advantage of
longitudinal CRIB in that high efficiency can be achieved without backward
retrieval. Besides unlike CRIB, this scheme does not require a direct control of atomic levels, thus potentially reduces decoherence in the system. The scheme can be implemented in resonant media which do not demonstrate linear Stark or Zeeman effects and allows one to combine Raman-interaction-based and CRIB-based approaches solely in the framework of the former.

\begin{acknowledgments}
This research was supported by NSF (Grant No. $0855688$) and RFBR (Grant No. 12-02-00651).
\end{acknowledgments}

{} {} {}
\bibliography{BibTeXfile}

\begin{thebibliography}{47}%
\makeatletter
\providecommand \@ifxundefined [1]{%
 \@ifx{#1\undefined}
}%
\providecommand \@ifnum [1]{%
 \ifnum #1\expandafter \@firstoftwo
 \else \expandafter \@secondoftwo
 \fi
}%
\providecommand \@ifx [1]{%
 \ifx #1\expandafter \@firstoftwo
 \else \expandafter \@secondoftwo
 \fi
}%
\providecommand \natexlab [1]{#1}%
\providecommand \enquote  [1]{``#1''}%
\providecommand \bibnamefont  [1]{#1}%
\providecommand \bibfnamefont [1]{#1}%
\providecommand \citenamefont [1]{#1}%
\providecommand \href@noop [0]{\@secondoftwo}%
\providecommand \href [0]{\begingroup \@sanitize@url \@href}%
\providecommand \@href[1]{\@@startlink{#1}\@@href}%
\providecommand \@@href[1]{\endgroup#1\@@endlink}%
\providecommand \@sanitize@url [0]{\catcode `\\12\catcode `\$12\catcode
  `\&12\catcode `\#12\catcode `\^12\catcode `\_12\catcode `\%12\relax}%
\providecommand \@@startlink[1]{}%
\providecommand \@@endlink[0]{}%
\providecommand \url  [0]{\begingroup\@sanitize@url \@url }%
\providecommand \@url [1]{\endgroup\@href {#1}{\urlprefix }}%
\providecommand \urlprefix  [0]{URL }%
\providecommand \Eprint [0]{\href }%
\providecommand \doibase [0]{http://dx.doi.org/}%
\providecommand \selectlanguage [0]{\@gobble}%
\providecommand \bibinfo  [0]{\@secondoftwo}%
\providecommand \bibfield  [0]{\@secondoftwo}%
\providecommand \translation [1]{[#1]}%
\providecommand \BibitemOpen [0]{}%
\providecommand \bibitemStop [0]{}%
\providecommand \bibitemNoStop [0]{.\EOS\space}%
\providecommand \EOS [0]{\spacefactor3000\relax}%
\providecommand \BibitemShut  [1]{\csname bibitem#1\endcsname}%
\let\auto@bib@innerbib\@empty
\bibitem [{\citenamefont {Hammerer}\ \emph {et~al.}(2010)\citenamefont
  {Hammerer}, \citenamefont {S{\o}rensen},\ and\ \citenamefont
  {Polzik}}]{HSP_2010}%
  \BibitemOpen
  \bibfield  {author} {\bibinfo {author} {\bibfnamefont {K.}~\bibnamefont
  {Hammerer}}, \bibinfo {author} {\bibfnamefont {A.~S.}\ \bibnamefont
  {S{\o}rensen}}, \ and\ \bibinfo {author} {\bibfnamefont {E.~S.}\ \bibnamefont
  {Polzik}},\ }\href@noop {} {\bibfield  {journal} {\bibinfo  {journal} {Rev.\
  Mod.\ Phys.}\ }\textbf {\bibinfo {volume} {82}},\ \bibinfo {pages} {1041}
  (\bibinfo {year} {2010})}\BibitemShut {NoStop}%
\bibitem [{\citenamefont {Tittel}\ \emph {et~al.}(2010)\citenamefont {Tittel},
  \citenamefont {Afzelius}, \citenamefont {Cone}, \citenamefont
  {Chaneli{\`{e}}re}, \citenamefont {Kr{\"{o}}ll}, \citenamefont {Moiseev},\
  and\ \citenamefont {Sellars}}]{TACCKMS_2010}%
  \BibitemOpen
  \bibfield  {author} {\bibinfo {author} {\bibfnamefont {W.}~\bibnamefont
  {Tittel}}, \bibinfo {author} {\bibfnamefont {M.}~\bibnamefont {Afzelius}},
  \bibinfo {author} {\bibfnamefont {R.~L.}\ \bibnamefont {Cone}}, \bibinfo
  {author} {\bibfnamefont {T.}~\bibnamefont {Chaneli{\`{e}}re}}, \bibinfo
  {author} {\bibfnamefont {S.}~\bibnamefont {Kr{\"{o}}ll}}, \bibinfo {author}
  {\bibfnamefont {S.~A.}\ \bibnamefont {Moiseev}}, \ and\ \bibinfo {author}
  {\bibfnamefont {M.}~\bibnamefont {Sellars}},\ }\href@noop {} {\bibfield
  {journal} {\bibinfo  {journal} {Laser \& Photonics Reviews}\ }\textbf
  {\bibinfo {volume} {4}},\ \bibinfo {pages} {244} (\bibinfo {year}
  {2010})}\BibitemShut {NoStop}%
\bibitem [{\citenamefont {Simon}\ \emph {et~al.}(2010)\citenamefont {Simon},
  \citenamefont {Afzelius}, \citenamefont {Appel}, \citenamefont {de~la
  Giroday}, \citenamefont {Dewhurst}, \citenamefont {Gisin}, \citenamefont
  {Hu}, \citenamefont {Jelezko}, \citenamefont {Kr{\"{o}}ll}, \citenamefont
  {M{\"{u}}ller}, \citenamefont {Nunn}, \citenamefont {Polzik}, \citenamefont
  {Rarity}, \citenamefont {Riedmatten}, \citenamefont {Rosenfeld},
  \citenamefont {Shields}, \citenamefont {Sk{\"{o}}ld}, \citenamefont
  {Stevenson}, \citenamefont {Thew}, \citenamefont {Walmsley}, \citenamefont
  {Weber}, \citenamefont {Weinfurter}, \citenamefont {J.Wrachtrup},\ and\
  \citenamefont {Young}}]{SAAB_2010}%
  \BibitemOpen
  \bibfield  {author} {\bibinfo {author} {\bibfnamefont {C.}~\bibnamefont
  {Simon}}, \bibinfo {author} {\bibfnamefont {M.}~\bibnamefont {Afzelius}},
  \bibinfo {author} {\bibfnamefont {J.}~\bibnamefont {Appel}}, \bibinfo
  {author} {\bibfnamefont {A.~B.}\ \bibnamefont {de~la Giroday}}, \bibinfo
  {author} {\bibfnamefont {S.}~\bibnamefont {Dewhurst}}, \bibinfo {author}
  {\bibfnamefont {N.}~\bibnamefont {Gisin}}, \bibinfo {author} {\bibfnamefont
  {C.}~\bibnamefont {Hu}}, \bibinfo {author} {\bibfnamefont {F.}~\bibnamefont
  {Jelezko}}, \bibinfo {author} {\bibfnamefont {S.}~\bibnamefont
  {Kr{\"{o}}ll}}, \bibinfo {author} {\bibfnamefont {J.}~\bibnamefont
  {M{\"{u}}ller}}, \bibinfo {author} {\bibfnamefont {J.}~\bibnamefont {Nunn}},
  \bibinfo {author} {\bibfnamefont {E.}~\bibnamefont {Polzik}}, \bibinfo
  {author} {\bibfnamefont {J.}~\bibnamefont {Rarity}}, \bibinfo {author}
  {\bibfnamefont {H.~D.}\ \bibnamefont {Riedmatten}}, \bibinfo {author}
  {\bibfnamefont {W.}~\bibnamefont {Rosenfeld}}, \bibinfo {author}
  {\bibfnamefont {A.}~\bibnamefont {Shields}}, \bibinfo {author} {\bibfnamefont
  {N.}~\bibnamefont {Sk{\"{o}}ld}}, \bibinfo {author} {\bibfnamefont
  {R.}~\bibnamefont {Stevenson}}, \bibinfo {author} {\bibfnamefont
  {R.}~\bibnamefont {Thew}}, \bibinfo {author} {\bibfnamefont {I.}~\bibnamefont
  {Walmsley}}, \bibinfo {author} {\bibfnamefont {M.}~\bibnamefont {Weber}},
  \bibinfo {author} {\bibfnamefont {H.}~\bibnamefont {Weinfurter}}, \bibinfo
  {author} {\bibnamefont {J.Wrachtrup}}, \ and\ \bibinfo {author}
  {\bibfnamefont {R.}~\bibnamefont {Young}},\ }\href@noop {} {\bibfield
  {journal} {\bibinfo  {journal} {The European Physical Journal D}\ }\textbf
  {\bibinfo {volume} {58}},\ \bibinfo {pages} {1} (\bibinfo {year}
  {2010})}\BibitemShut {NoStop}%
\bibitem [{\citenamefont {Kok}\ \emph {et~al.}(2007)\citenamefont {Kok},
  \citenamefont {Munro}, \citenamefont {Nemoto}, \citenamefont {Ralph},
  \citenamefont {Dowling},\ and\ \citenamefont {Milburn}}]{KMNRDM_2007}%
  \BibitemOpen
  \bibfield  {author} {\bibinfo {author} {\bibfnamefont {P.}~\bibnamefont
  {Kok}}, \bibinfo {author} {\bibfnamefont {W.~J.}\ \bibnamefont {Munro}},
  \bibinfo {author} {\bibfnamefont {K.}~\bibnamefont {Nemoto}}, \bibinfo
  {author} {\bibfnamefont {T.~C.}\ \bibnamefont {Ralph}}, \bibinfo {author}
  {\bibfnamefont {J.~P.}\ \bibnamefont {Dowling}}, \ and\ \bibinfo {author}
  {\bibfnamefont {G.~J.}\ \bibnamefont {Milburn}},\ }\href@noop {} {\bibfield
  {journal} {\bibinfo  {journal} {Rev.\ Mod.\ Phys.}\ }\textbf {\bibinfo
  {volume} {79}},\ \bibinfo {pages} {135} (\bibinfo {year} {2007})}\BibitemShut
  {NoStop}%
\bibitem [{\citenamefont {Sangouard}\ \emph {et~al.}(2011)\citenamefont
  {Sangouard}, \citenamefont {Simon}, \citenamefont {de~Riedmatten},\ and\
  \citenamefont {Gisin}}]{SSRG_2011}%
  \BibitemOpen
  \bibfield  {author} {\bibinfo {author} {\bibfnamefont {N.}~\bibnamefont
  {Sangouard}}, \bibinfo {author} {\bibfnamefont {C.}~\bibnamefont {Simon}},
  \bibinfo {author} {\bibfnamefont {H.}~\bibnamefont {de~Riedmatten}}, \ and\
  \bibinfo {author} {\bibfnamefont {N.}~\bibnamefont {Gisin}},\ }\href@noop {}
  {\bibfield  {journal} {\bibinfo  {journal} {Rev. Mod. Phys.}\ }\textbf
  {\bibinfo {volume} {83}},\ \bibinfo {pages} {33} (\bibinfo {year}
  {2011})}\BibitemShut {NoStop}%
\bibitem [{\citenamefont {Pittman}\ \emph {et~al.}(2002)\citenamefont
  {Pittman}, \citenamefont {Jacobs},\ and\ \citenamefont {Franson}}]{PJF_2002}%
  \BibitemOpen
  \bibfield  {author} {\bibinfo {author} {\bibfnamefont {T.~B.}\ \bibnamefont
  {Pittman}}, \bibinfo {author} {\bibfnamefont {B.~C.}\ \bibnamefont {Jacobs}},
  \ and\ \bibinfo {author} {\bibfnamefont {J.~D.}\ \bibnamefont {Franson}},\
  }\href@noop {} {\bibfield  {journal} {\bibinfo  {journal} {Phys.\ Rev.\ A}\
  }\textbf {\bibinfo {volume} {66}},\ \bibinfo {pages} {042303} (\bibinfo
  {year} {2002})}\BibitemShut {NoStop}%
\bibitem [{\citenamefont {Hosseini}\ \emph {et~al.}(2011)\citenamefont
  {Hosseini}, \citenamefont {Sparkes}, \citenamefont {Campbell}, \citenamefont
  {Lam},\ and\ \citenamefont {Buchler}}]{HSCLB_2011}%
  \BibitemOpen
  \bibfield  {author} {\bibinfo {author} {\bibfnamefont {M.}~\bibnamefont
  {Hosseini}}, \bibinfo {author} {\bibfnamefont {B.~M.}\ \bibnamefont
  {Sparkes}}, \bibinfo {author} {\bibfnamefont {G.}~\bibnamefont {Campbell}},
  \bibinfo {author} {\bibfnamefont {P.~K.}\ \bibnamefont {Lam}}, \ and\
  \bibinfo {author} {\bibfnamefont {B.~C.}\ \bibnamefont {Buchler}},\
  }\href@noop {} {\bibfield  {journal} {\bibinfo  {journal} {Nature
  Communications}\ }\textbf {\bibinfo {volume} {2}},\ \bibinfo {pages} {174}
  (\bibinfo {year} {2011})}\BibitemShut {NoStop}%
\bibitem [{\citenamefont {Hedges}\ \emph {et~al.}(2010)\citenamefont {Hedges},
  \citenamefont {Longdell}, \citenamefont {Li},\ and\ \citenamefont
  {Sellars}}]{HLLS_2010}%
  \BibitemOpen
  \bibfield  {author} {\bibinfo {author} {\bibfnamefont {M.~P.}\ \bibnamefont
  {Hedges}}, \bibinfo {author} {\bibfnamefont {J.~J.}\ \bibnamefont
  {Longdell}}, \bibinfo {author} {\bibfnamefont {Y.}~\bibnamefont {Li}}, \ and\
  \bibinfo {author} {\bibfnamefont {M.~J.}\ \bibnamefont {Sellars}},\
  }\href@noop {} {\bibfield  {journal} {\bibinfo  {journal} {Nature (London)}\
  }\textbf {\bibinfo {volume} {465}},\ \bibinfo {pages} {1052} (\bibinfo {year}
  {2010})}\BibitemShut {NoStop}%
\bibitem [{\citenamefont {Chaneli{\`{e}}re}\ \emph {et~al.}(2010)\citenamefont
  {Chaneli{\`{e}}re}, \citenamefont {Ruggiero}, \citenamefont {Bonarota},
  \citenamefont {Afzelius},\ and\ \citenamefont {Gou{\"{e}}t}}]{CRBAG_2010}%
  \BibitemOpen
  \bibfield  {author} {\bibinfo {author} {\bibfnamefont {T.}~\bibnamefont
  {Chaneli{\`{e}}re}}, \bibinfo {author} {\bibfnamefont {J.}~\bibnamefont
  {Ruggiero}}, \bibinfo {author} {\bibfnamefont {M.}~\bibnamefont {Bonarota}},
  \bibinfo {author} {\bibfnamefont {M.}~\bibnamefont {Afzelius}}, \ and\
  \bibinfo {author} {\bibfnamefont {J.-L.~L.}\ \bibnamefont {Gou{\"{e}}t}},\
  }\href@noop {} {\bibfield  {journal} {\bibinfo  {journal} {New Journal of
  Physics}\ }\textbf {\bibinfo {volume} {12}} (\bibinfo {year}
  {2010})}\BibitemShut {NoStop}%
\bibitem [{\citenamefont {Sabooni}\ \emph {et~al.}(2010)\citenamefont
  {Sabooni}, \citenamefont {Beaudoin}, \citenamefont {Walther}, \citenamefont
  {Lin}, \citenamefont {Amari}, \citenamefont {Huang},\ and\ \citenamefont
  {Kr{\"{o}}ll}}]{SBWLAHK_2010}%
  \BibitemOpen
  \bibfield  {author} {\bibinfo {author} {\bibfnamefont {M.}~\bibnamefont
  {Sabooni}}, \bibinfo {author} {\bibfnamefont {F.}~\bibnamefont {Beaudoin}},
  \bibinfo {author} {\bibfnamefont {A.}~\bibnamefont {Walther}}, \bibinfo
  {author} {\bibfnamefont {N.}~\bibnamefont {Lin}}, \bibinfo {author}
  {\bibfnamefont {A.}~\bibnamefont {Amari}}, \bibinfo {author} {\bibfnamefont
  {M.}~\bibnamefont {Huang}}, \ and\ \bibinfo {author} {\bibfnamefont
  {S.}~\bibnamefont {Kr{\"{o}}ll}},\ }\href@noop {} {\bibfield  {journal}
  {\bibinfo  {journal} {Phys.\ Rev.\ Lett.}\ }\textbf {\bibinfo {volume}
  {105}},\ \bibinfo {pages} {060501} (\bibinfo {year} {2010})}\BibitemShut
  {NoStop}%
\bibitem [{\citenamefont {Afzelius}\ \emph {et~al.}(2010)\citenamefont
  {Afzelius}, \citenamefont {Usmani}, \citenamefont {Amari}, \citenamefont
  {Lauritzen}, \citenamefont {Walther}, \citenamefont {Simon}, \citenamefont
  {Sangouard}, \citenamefont {Min{\'{a}}{\v{r}}}, \citenamefont
  {de~Riedmatten}, \citenamefont {Gisin},\ and\ \citenamefont
  {Kr{\"{o}}ll}}]{AUALWSSRGK_2010}%
  \BibitemOpen
  \bibfield  {author} {\bibinfo {author} {\bibfnamefont {M.}~\bibnamefont
  {Afzelius}}, \bibinfo {author} {\bibfnamefont {I.}~\bibnamefont {Usmani}},
  \bibinfo {author} {\bibfnamefont {A.}~\bibnamefont {Amari}}, \bibinfo
  {author} {\bibfnamefont {B.}~\bibnamefont {Lauritzen}}, \bibinfo {author}
  {\bibfnamefont {A.}~\bibnamefont {Walther}}, \bibinfo {author} {\bibfnamefont
  {C.}~\bibnamefont {Simon}}, \bibinfo {author} {\bibfnamefont
  {N.}~\bibnamefont {Sangouard}}, \bibinfo {author} {\bibfnamefont
  {J.}~\bibnamefont {Min{\'{a}}{\v{r}}}}, \bibinfo {author} {\bibfnamefont
  {H.}~\bibnamefont {de~Riedmatten}}, \bibinfo {author} {\bibfnamefont
  {N.}~\bibnamefont {Gisin}}, \ and\ \bibinfo {author} {\bibfnamefont
  {S.}~\bibnamefont {Kr{\"{o}}ll}},\ }\href@noop {} {\bibfield  {journal}
  {\bibinfo  {journal} {Phys.\ Rev.\ Lett.}\ }\textbf {\bibinfo {volume}
  {104}},\ \bibinfo {pages} {040503} (\bibinfo {year} {2010})}\BibitemShut
  {NoStop}%
\bibitem [{\citenamefont {Usmani}\ \emph {et~al.}(2010)\citenamefont {Usmani},
  \citenamefont {Afzelius}, \citenamefont {de~Riedmatten},\ and\ \citenamefont
  {Gisin}}]{UARG_2010}%
  \BibitemOpen
  \bibfield  {author} {\bibinfo {author} {\bibfnamefont {I.}~\bibnamefont
  {Usmani}}, \bibinfo {author} {\bibfnamefont {M.}~\bibnamefont {Afzelius}},
  \bibinfo {author} {\bibfnamefont {H.}~\bibnamefont {de~Riedmatten}}, \ and\
  \bibinfo {author} {\bibfnamefont {N.}~\bibnamefont {Gisin}},\ }\href@noop {}
  {\bibfield  {journal} {\bibinfo  {journal} {Nature Communications}\ }\textbf
  {\bibinfo {volume} {1}},\ \bibinfo {pages} {1} (\bibinfo {year}
  {2010})}\BibitemShut {NoStop}%
\bibitem [{\citenamefont {Saglamyurek}\ \emph {et~al.}(2011)\citenamefont
  {Saglamyurek}, \citenamefont {Sinclair}, \citenamefont {Jin}, \citenamefont
  {Slater}, \citenamefont {Oblak}, \citenamefont {Bussi{\`{e}}res},
  \citenamefont {George}, \citenamefont {Ricken}, \citenamefont {Sohler},\ and\
  \citenamefont {Tittel}}]{SSJSOBGRST_2011}%
  \BibitemOpen
  \bibfield  {author} {\bibinfo {author} {\bibfnamefont {E.}~\bibnamefont
  {Saglamyurek}}, \bibinfo {author} {\bibfnamefont {N.}~\bibnamefont
  {Sinclair}}, \bibinfo {author} {\bibfnamefont {J.}~\bibnamefont {Jin}},
  \bibinfo {author} {\bibfnamefont {J.~A.}\ \bibnamefont {Slater}}, \bibinfo
  {author} {\bibfnamefont {D.}~\bibnamefont {Oblak}}, \bibinfo {author}
  {\bibfnamefont {F.}~\bibnamefont {Bussi{\`{e}}res}}, \bibinfo {author}
  {\bibfnamefont {M.}~\bibnamefont {George}}, \bibinfo {author} {\bibfnamefont
  {R.}~\bibnamefont {Ricken}}, \bibinfo {author} {\bibfnamefont
  {W.}~\bibnamefont {Sohler}}, \ and\ \bibinfo {author} {\bibfnamefont
  {W.}~\bibnamefont {Tittel}},\ }\href@noop {} {\bibfield  {journal} {\bibinfo
  {journal} {Nature}\ }\textbf {\bibinfo {volume} {469}},\ \bibinfo {pages}
  {512} (\bibinfo {year} {2011})}\BibitemShut {NoStop}%
\bibitem [{\citenamefont {Clausen}\ \emph {et~al.}(2011)\citenamefont
  {Clausen}, \citenamefont {Usmani}, \citenamefont {Bussi{\`{e}}res},
  \citenamefont {Sangouard}, \citenamefont {Afzelius}, \citenamefont
  {de~Riedmatten},\ and\ \citenamefont {Gisin}}]{CUBSARG_2011}%
  \BibitemOpen
  \bibfield  {author} {\bibinfo {author} {\bibfnamefont {C.}~\bibnamefont
  {Clausen}}, \bibinfo {author} {\bibfnamefont {I.}~\bibnamefont {Usmani}},
  \bibinfo {author} {\bibfnamefont {F.}~\bibnamefont {Bussi{\`{e}}res}},
  \bibinfo {author} {\bibfnamefont {N.}~\bibnamefont {Sangouard}}, \bibinfo
  {author} {\bibfnamefont {M.}~\bibnamefont {Afzelius}}, \bibinfo {author}
  {\bibfnamefont {H.}~\bibnamefont {de~Riedmatten}}, \ and\ \bibinfo {author}
  {\bibfnamefont {N.}~\bibnamefont {Gisin}},\ }\href@noop {} {\bibfield
  {journal} {\bibinfo  {journal} {Nature}\ }\textbf {\bibinfo {volume} {469}},\
  \bibinfo {pages} {508} (\bibinfo {year} {2011})}\BibitemShut {NoStop}%
\bibitem [{\citenamefont {Chaneli\`{e}re}\ \emph {et~al.}(2005)\citenamefont
  {Chaneli\`{e}re}, \citenamefont {Matsukevich}, \citenamefont {Jenkins},
  \citenamefont {Lan}, \citenamefont {Kennedy},\ and\ \citenamefont
  {Kuzmich}}]{Chaneliere05}%
  \BibitemOpen
  \bibfield  {author} {\bibinfo {author} {\bibfnamefont {T.}~\bibnamefont
  {Chaneli\`{e}re}}, \bibinfo {author} {\bibfnamefont {D.~N.}\ \bibnamefont
  {Matsukevich}}, \bibinfo {author} {\bibfnamefont {S.~D.}\ \bibnamefont
  {Jenkins}}, \bibinfo {author} {\bibfnamefont {S.~Y.}\ \bibnamefont {Lan}},
  \bibinfo {author} {\bibfnamefont {T.~A.~B.}\ \bibnamefont {Kennedy}}, \ and\
  \bibinfo {author} {\bibfnamefont {A.}~\bibnamefont {Kuzmich}},\ }\href@noop
  {} {\bibfield  {journal} {\bibinfo  {journal} {Nature}\ }\textbf {\bibinfo
  {volume} {438}},\ \bibinfo {pages} {833} (\bibinfo {year}
  {2005})}\BibitemShut {NoStop}%
\bibitem [{\citenamefont {Eisaman}\ \emph {et~al.}(2005)\citenamefont
  {Eisaman}, \citenamefont {Andre}, \citenamefont {Massou}, \citenamefont
  {Fleischhauer}, \citenamefont {Zibrov},\ and\ \citenamefont
  {Lukin}}]{Eisaman05}%
  \BibitemOpen
  \bibfield  {author} {\bibinfo {author} {\bibfnamefont {M.~D.}\ \bibnamefont
  {Eisaman}}, \bibinfo {author} {\bibfnamefont {A.}~\bibnamefont {Andre}},
  \bibinfo {author} {\bibfnamefont {F.}~\bibnamefont {Massou}}, \bibinfo
  {author} {\bibfnamefont {M.}~\bibnamefont {Fleischhauer}}, \bibinfo {author}
  {\bibfnamefont {A.~S.}\ \bibnamefont {Zibrov}}, \ and\ \bibinfo {author}
  {\bibfnamefont {M.~D.}\ \bibnamefont {Lukin}},\ }\href@noop {} {\bibfield
  {journal} {\bibinfo  {journal} {Nature}\ }\textbf {\bibinfo {volume} {438}},\
  \bibinfo {pages} {837} (\bibinfo {year} {2005})}\BibitemShut {NoStop}%
\bibitem [{\citenamefont {Novikova}\ \emph {et~al.}(2007)\citenamefont
  {Novikova}, \citenamefont {Gorshkov}, \citenamefont {Phillips}, \citenamefont
  {S{\o}rensen}, \citenamefont {Lukin},\ and\ \citenamefont
  {Walsworth}}]{Novikova07}%
  \BibitemOpen
  \bibfield  {author} {\bibinfo {author} {\bibfnamefont {I.}~\bibnamefont
  {Novikova}}, \bibinfo {author} {\bibfnamefont {A.~V.}\ \bibnamefont
  {Gorshkov}}, \bibinfo {author} {\bibfnamefont {D.~F.}\ \bibnamefont
  {Phillips}}, \bibinfo {author} {\bibfnamefont {A.~S.}\ \bibnamefont
  {S{\o}rensen}}, \bibinfo {author} {\bibfnamefont {M.~D.}\ \bibnamefont
  {Lukin}}, \ and\ \bibinfo {author} {\bibfnamefont {R.~L.}\ \bibnamefont
  {Walsworth}},\ }\href@noop {} {\bibfield  {journal} {\bibinfo  {journal}
  {Physical Review Letters}\ }\textbf {\bibinfo {volume} {98}},\ \bibinfo
  {pages} {243602} (\bibinfo {year} {2007})}\BibitemShut {NoStop}%
\bibitem [{\citenamefont {Choi}\ \emph {et~al.}(2008)\citenamefont {Choi},
  \citenamefont {Deng}, \citenamefont {Laurat},\ and\ \citenamefont
  {Kimble}}]{Choi08}%
  \BibitemOpen
  \bibfield  {author} {\bibinfo {author} {\bibfnamefont {K.~S.}\ \bibnamefont
  {Choi}}, \bibinfo {author} {\bibfnamefont {H.}~\bibnamefont {Deng}}, \bibinfo
  {author} {\bibfnamefont {J.}~\bibnamefont {Laurat}}, \ and\ \bibinfo {author}
  {\bibfnamefont {H.~J.}\ \bibnamefont {Kimble}},\ }\href@noop {} {\bibfield
  {journal} {\bibinfo  {journal} {Nature}\ }\textbf {\bibinfo {volume} {452}},\
  \bibinfo {pages} {67} (\bibinfo {year} {2008})}\BibitemShut {NoStop}%
\bibitem [{\citenamefont {Reim}\ \emph {et~al.}(2010)\citenamefont {Reim},
  \citenamefont {Nunn}, \citenamefont {Lorenz}, \citenamefont {Sussman},
  \citenamefont {Lee}, \citenamefont {Langford}, \citenamefont {Jaksch},\ and\
  \citenamefont {Walmsley}}]{RNLSLLJW_2010}%
  \BibitemOpen
  \bibfield  {author} {\bibinfo {author} {\bibfnamefont {K.~F.}\ \bibnamefont
  {Reim}}, \bibinfo {author} {\bibfnamefont {J.}~\bibnamefont {Nunn}}, \bibinfo
  {author} {\bibfnamefont {V.~O.}\ \bibnamefont {Lorenz}}, \bibinfo {author}
  {\bibfnamefont {B.~J.}\ \bibnamefont {Sussman}}, \bibinfo {author}
  {\bibfnamefont {K.~C.}\ \bibnamefont {Lee}}, \bibinfo {author} {\bibfnamefont
  {N.~K.}\ \bibnamefont {Langford}}, \bibinfo {author} {\bibfnamefont
  {D.}~\bibnamefont {Jaksch}}, \ and\ \bibinfo {author} {\bibfnamefont {I.~A.}\
  \bibnamefont {Walmsley}},\ }\href@noop {} {\bibfield  {journal} {\bibinfo
  {journal} {Nature Photonics}\ }\textbf {\bibinfo {volume} {4}},\ \bibinfo
  {pages} {218} (\bibinfo {year} {2010})}\BibitemShut {NoStop}%
\bibitem [{\citenamefont {Reim}\ \emph {et~al.}(2011)\citenamefont {Reim},
  \citenamefont {Michelberger}, \citenamefont {Lee}, \citenamefont {Nunn},
  \citenamefont {Langford},\ and\ \citenamefont {Walmsley}}]{RMLNLW_2011}%
  \BibitemOpen
  \bibfield  {author} {\bibinfo {author} {\bibfnamefont {K.~F.}\ \bibnamefont
  {Reim}}, \bibinfo {author} {\bibfnamefont {P.}~\bibnamefont {Michelberger}},
  \bibinfo {author} {\bibfnamefont {K.~C.}\ \bibnamefont {Lee}}, \bibinfo
  {author} {\bibfnamefont {J.}~\bibnamefont {Nunn}}, \bibinfo {author}
  {\bibfnamefont {N.~K.}\ \bibnamefont {Langford}}, \ and\ \bibinfo {author}
  {\bibfnamefont {I.~A.}\ \bibnamefont {Walmsley}},\ }\href@noop {} {\bibfield
  {journal} {\bibinfo  {journal} {Phys. Rev. Lett.}\ }\textbf {\bibinfo
  {volume} {107}},\ \bibinfo {pages} {053603} (\bibinfo {year}
  {2011})}\BibitemShut {NoStop}%
\bibitem [{\citenamefont {Kalachev}\ and\ \citenamefont
  {Kocharovskaya}(2011{\natexlab{a}})}]{Kalachev11}%
  \BibitemOpen
  \bibfield  {author} {\bibinfo {author} {\bibfnamefont {A.}~\bibnamefont
  {Kalachev}}\ and\ \bibinfo {author} {\bibfnamefont {O.}~\bibnamefont
  {Kocharovskaya}},\ }\href@noop {} {\bibfield  {journal} {\bibinfo  {journal}
  {Physical Review A}\ }\textbf {\bibinfo {volume} {83}},\ \bibinfo {pages}
  {053849} (\bibinfo {year} {2011}{\natexlab{a}})}\BibitemShut {NoStop}%
\bibitem [{Zha()}]{ZhangTBP}%
  \BibitemOpen
  \href@noop {} {}\bibinfo {note} {X. Zhang, A. Kalachev, and O. Kocharovskaya,
  to be published}\BibitemShut {NoStop}%
\bibitem [{\citenamefont {Clark}\ \emph {et~al.}(2012)\citenamefont {Clark},
  \citenamefont {Heshami},\ and\ \citenamefont {Simon}}]{Clark12}%
  \BibitemOpen
  \bibfield  {author} {\bibinfo {author} {\bibfnamefont {J.}~\bibnamefont
  {Clark}}, \bibinfo {author} {\bibfnamefont {K.}~\bibnamefont {Heshami}}, \
  and\ \bibinfo {author} {\bibfnamefont {C.}~\bibnamefont {Simon}},\
  }\href@noop {} {\bibfield  {journal} {\bibinfo  {journal} {Physical Review
  A}\ }\textbf {\bibinfo {volume} {86}},\ \bibinfo {pages} {013833} (\bibinfo
  {year} {2012})}\BibitemShut {NoStop}%
\bibitem [{\citenamefont {Kalachev}\ and\ \citenamefont
  {Kocharovskaya}(2011{\natexlab{b}})}]{Kalachev11JMO}%
  \BibitemOpen
  \bibfield  {author} {\bibinfo {author} {\bibfnamefont {A.}~\bibnamefont
  {Kalachev}}\ and\ \bibinfo {author} {\bibfnamefont {O.}~\bibnamefont
  {Kocharovskaya}},\ }\href@noop {} {\bibfield  {journal} {\bibinfo  {journal}
  {Journal of Modern Optics}\ }\textbf {\bibinfo {volume} {58}},\ \bibinfo
  {pages} {1971} (\bibinfo {year} {2011}{\natexlab{b}})}\BibitemShut {NoStop}%
\bibitem [{\citenamefont {Milonni}(1995)}]{Milonni95}%
  \BibitemOpen
  \bibfield  {author} {\bibinfo {author} {\bibfnamefont {P.~W.}\ \bibnamefont
  {Milonni}},\ }\href@noop {} {\bibfield  {journal} {\bibinfo  {journal}
  {Journal of Modern Optics}\ }\textbf {\bibinfo {volume} {42}},\ \bibinfo
  {pages} {1991} (\bibinfo {year} {1995})}\BibitemShut {NoStop}%
\bibitem [{\citenamefont {Alexander}\ \emph {et~al.}(2006)\citenamefont
  {Alexander}, \citenamefont {Longdell}, \citenamefont {Sellars},\ and\
  \citenamefont {Manson}}]{Alexander06}%
  \BibitemOpen
  \bibfield  {author} {\bibinfo {author} {\bibfnamefont {A.~L.}\ \bibnamefont
  {Alexander}}, \bibinfo {author} {\bibfnamefont {J.~J.}\ \bibnamefont
  {Longdell}}, \bibinfo {author} {\bibfnamefont {M.~J.}\ \bibnamefont
  {Sellars}}, \ and\ \bibinfo {author} {\bibfnamefont {N.~B.}\ \bibnamefont
  {Manson}},\ }\href@noop {} {\bibfield  {journal} {\bibinfo  {journal}
  {Physical Review Letters}\ }\textbf {\bibinfo {volume} {96}},\ \bibinfo
  {pages} {043602} (\bibinfo {year} {2006})}\BibitemShut {NoStop}%
\bibitem [{\citenamefont {Longdell}\ \emph {et~al.}(2008)\citenamefont
  {Longdell}, \citenamefont {H\'{e}tet}, \citenamefont {Lam},\ and\
  \citenamefont {Sellars}}]{Longdell08}%
  \BibitemOpen
  \bibfield  {author} {\bibinfo {author} {\bibfnamefont {J.~J.}\ \bibnamefont
  {Longdell}}, \bibinfo {author} {\bibfnamefont {G.}~\bibnamefont {H\'{e}tet}},
  \bibinfo {author} {\bibfnamefont {P.~K.}\ \bibnamefont {Lam}}, \ and\
  \bibinfo {author} {\bibfnamefont {M.~J.}\ \bibnamefont {Sellars}},\
  }\href@noop {} {\bibfield  {journal} {\bibinfo  {journal} {Physical Review
  A}\ }\textbf {\bibinfo {volume} {78}},\ \bibinfo {pages} {032337} (\bibinfo
  {year} {2008})}\BibitemShut {NoStop}%
\bibitem [{\citenamefont {Moiseev}\ and\ \citenamefont
  {Arslanov}(2008)}]{Moiseev08}%
  \BibitemOpen
  \bibfield  {author} {\bibinfo {author} {\bibfnamefont {S.~A.}\ \bibnamefont
  {Moiseev}}\ and\ \bibinfo {author} {\bibfnamefont {N.~M.}\ \bibnamefont
  {Arslanov}},\ }\href@noop {} {\bibfield  {journal} {\bibinfo  {journal}
  {Physical Review A}\ }\textbf {\bibinfo {volume} {78}},\ \bibinfo {pages}
  {023803} (\bibinfo {year} {2008})}\BibitemShut {NoStop}%
\bibitem [{\citenamefont {H\'{e}tet}\ \emph
  {et~al.}(2008{\natexlab{a}})\citenamefont {H\'{e}tet}, \citenamefont
  {Longdell}, \citenamefont {Sellars}, \citenamefont {Lam},\ and\ \citenamefont
  {Buchler}}]{Hetet08}%
  \BibitemOpen
  \bibfield  {author} {\bibinfo {author} {\bibfnamefont {G.}~\bibnamefont
  {H\'{e}tet}}, \bibinfo {author} {\bibfnamefont {J.~J.}\ \bibnamefont
  {Longdell}}, \bibinfo {author} {\bibfnamefont {M.~J.}\ \bibnamefont
  {Sellars}}, \bibinfo {author} {\bibfnamefont {P.~K.}\ \bibnamefont {Lam}}, \
  and\ \bibinfo {author} {\bibfnamefont {B.~C.}\ \bibnamefont {Buchler}},\
  }\href@noop {} {\bibfield  {journal} {\bibinfo  {journal} {Physical Review
  Letters}\ }\textbf {\bibinfo {volume} {101}},\ \bibinfo {pages} {203601}
  (\bibinfo {year} {2008}{\natexlab{a}})}\BibitemShut {NoStop}%
\bibitem [{\citenamefont {H\'{e}tet}\ \emph
  {et~al.}(2008{\natexlab{b}})\citenamefont {H\'{e}tet}, \citenamefont
  {Longdell}, \citenamefont {Alexander}, \citenamefont {Lam},\ and\
  \citenamefont {Sellars}}]{Hetet08EXP}%
  \BibitemOpen
  \bibfield  {author} {\bibinfo {author} {\bibfnamefont {G.}~\bibnamefont
  {H\'{e}tet}}, \bibinfo {author} {\bibfnamefont {J.~J.}\ \bibnamefont
  {Longdell}}, \bibinfo {author} {\bibfnamefont {A.~L.}\ \bibnamefont
  {Alexander}}, \bibinfo {author} {\bibfnamefont {P.~K.}\ \bibnamefont {Lam}},
  \ and\ \bibinfo {author} {\bibfnamefont {M.~J.}\ \bibnamefont {Sellars}},\
  }\href@noop {} {\bibfield  {journal} {\bibinfo  {journal} {Physical Review
  Letters}\ }\textbf {\bibinfo {volume} {100}},\ \bibinfo {pages} {023601}
  (\bibinfo {year} {2008}{\natexlab{b}})}\BibitemShut {NoStop}%
\bibitem [{\citenamefont {H\'{e}tet}\ \emph
  {et~al.}(2008{\natexlab{c}})\citenamefont {H\'{e}tet}, \citenamefont
  {Hosseini}, \citenamefont {Sparkes}, \citenamefont {Oblak}, \citenamefont
  {Lam},\ and\ \citenamefont {Buchler}}]{Hetet08EXP3level}%
  \BibitemOpen
  \bibfield  {author} {\bibinfo {author} {\bibfnamefont {G.}~\bibnamefont
  {H\'{e}tet}}, \bibinfo {author} {\bibfnamefont {M.}~\bibnamefont {Hosseini}},
  \bibinfo {author} {\bibfnamefont {B.~M.}\ \bibnamefont {Sparkes}}, \bibinfo
  {author} {\bibfnamefont {D.}~\bibnamefont {Oblak}}, \bibinfo {author}
  {\bibfnamefont {P.~K.}\ \bibnamefont {Lam}}, \ and\ \bibinfo {author}
  {\bibfnamefont {B.~C.}\ \bibnamefont {Buchler}},\ }\href@noop {} {\bibfield
  {journal} {\bibinfo  {journal} {Optics Letters}\ }\textbf {\bibinfo {volume}
  {33}},\ \bibinfo {pages} {2323} (\bibinfo {year}
  {2008}{\natexlab{c}})}\BibitemShut {NoStop}%
\bibitem [{\citenamefont {Moiseev}\ and\ \citenamefont
  {Kr\"{o}ll}(2001)}]{Moiseev01}%
  \BibitemOpen
  \bibfield  {author} {\bibinfo {author} {\bibfnamefont {S.~A.}\ \bibnamefont
  {Moiseev}}\ and\ \bibinfo {author} {\bibfnamefont {S.}~\bibnamefont
  {Kr\"{o}ll}},\ }\href@noop {} {\bibfield  {journal} {\bibinfo  {journal}
  {Physical Review Letters}\ }\textbf {\bibinfo {volume} {87}},\ \bibinfo
  {pages} {173601} (\bibinfo {year} {2001})}\BibitemShut {NoStop}%
\bibitem [{\citenamefont {Nilsson}\ and\ \citenamefont
  {Kr\"{o}ll}(2005)}]{Nilsson05}%
  \BibitemOpen
  \bibfield  {author} {\bibinfo {author} {\bibfnamefont {M.}~\bibnamefont
  {Nilsson}}\ and\ \bibinfo {author} {\bibfnamefont {S.}~\bibnamefont
  {Kr\"{o}ll}},\ }\href@noop {} {\bibfield  {journal} {\bibinfo  {journal}
  {Optics Communications}\ }\textbf {\bibinfo {volume} {247}},\ \bibinfo
  {pages} {393} (\bibinfo {year} {2005})}\BibitemShut {NoStop}%
\bibitem [{\citenamefont {Surmacz}\ \emph {et~al.}(2008)\citenamefont
  {Surmacz}, \citenamefont {Nunn}, \citenamefont {Reim}, \citenamefont {Lee},
  \citenamefont {Lorenz}, \citenamefont {Sussman}, \citenamefont {Walmsley},\
  and\ \citenamefont {Jaksch}}]{Surmacz08}%
  \BibitemOpen
  \bibfield  {author} {\bibinfo {author} {\bibfnamefont {K.}~\bibnamefont
  {Surmacz}}, \bibinfo {author} {\bibfnamefont {J.}~\bibnamefont {Nunn}},
  \bibinfo {author} {\bibfnamefont {K.}~\bibnamefont {Reim}}, \bibinfo {author}
  {\bibfnamefont {K.~C.}\ \bibnamefont {Lee}}, \bibinfo {author} {\bibfnamefont
  {V.~O.}\ \bibnamefont {Lorenz}}, \bibinfo {author} {\bibfnamefont
  {B.}~\bibnamefont {Sussman}}, \bibinfo {author} {\bibfnamefont {I.~A.}\
  \bibnamefont {Walmsley}}, \ and\ \bibinfo {author} {\bibfnamefont
  {D.}~\bibnamefont {Jaksch}},\ }\href@noop {} {\bibfield  {journal} {\bibinfo
  {journal} {Physical Review A}\ }\textbf {\bibinfo {volume} {78}},\ \bibinfo
  {pages} {033806} (\bibinfo {year} {2008})}\BibitemShut {NoStop}%
\bibitem [{\citenamefont {Buchler}\ \emph {et~al.}(2010)\citenamefont
  {Buchler}, \citenamefont {Hosseini}, \citenamefont {H\'{e}tet}, \citenamefont
  {Sparkes},\ and\ \citenamefont {Lam}}]{Buchler11}%
  \BibitemOpen
  \bibfield  {author} {\bibinfo {author} {\bibfnamefont {B.~C.}\ \bibnamefont
  {Buchler}}, \bibinfo {author} {\bibfnamefont {M.}~\bibnamefont {Hosseini}},
  \bibinfo {author} {\bibfnamefont {G.}~\bibnamefont {H\'{e}tet}}, \bibinfo
  {author} {\bibfnamefont {B.~M.}\ \bibnamefont {Sparkes}}, \ and\ \bibinfo
  {author} {\bibfnamefont {P.~K.}\ \bibnamefont {Lam}},\ }\href@noop {}
  {\bibfield  {journal} {\bibinfo  {journal} {Optics Letters}\ }\textbf
  {\bibinfo {volume} {35}},\ \bibinfo {pages} {1091} (\bibinfo {year}
  {2010})}\BibitemShut {NoStop}%
\bibitem [{\citenamefont {Schnorrberger}\ \emph {et~al.}(2009)\citenamefont
  {Schnorrberger}, \citenamefont {Thompson}, \citenamefont {Trotzky},
  \citenamefont {Pugatch}, \citenamefont {Davidson}, \citenamefont {Kuhr},\
  and\ \citenamefont {Bloch}}]{STTPDKB_2009}%
  \BibitemOpen
  \bibfield  {author} {\bibinfo {author} {\bibfnamefont {U.}~\bibnamefont
  {Schnorrberger}}, \bibinfo {author} {\bibfnamefont {J.~D.}\ \bibnamefont
  {Thompson}}, \bibinfo {author} {\bibfnamefont {S.}~\bibnamefont {Trotzky}},
  \bibinfo {author} {\bibfnamefont {R.}~\bibnamefont {Pugatch}}, \bibinfo
  {author} {\bibfnamefont {N.}~\bibnamefont {Davidson}}, \bibinfo {author}
  {\bibfnamefont {S.}~\bibnamefont {Kuhr}}, \ and\ \bibinfo {author}
  {\bibfnamefont {I.}~\bibnamefont {Bloch}},\ }\href@noop {} {\bibfield
  {journal} {\bibinfo  {journal} {Phys. Rev. Lett.}\ }\textbf {\bibinfo
  {volume} {103}},\ \bibinfo {pages} {033003} (\bibinfo {year}
  {2009})}\BibitemShut {NoStop}%
\bibitem [{\citenamefont {Dudin}\ \emph {et~al.}(2010)\citenamefont {Dudin},
  \citenamefont {Zhao}, \citenamefont {Kennedy},\ and\ \citenamefont
  {Kuzmich}}]{DZKK_2010}%
  \BibitemOpen
  \bibfield  {author} {\bibinfo {author} {\bibfnamefont {Y.~O.}\ \bibnamefont
  {Dudin}}, \bibinfo {author} {\bibfnamefont {R.}~\bibnamefont {Zhao}},
  \bibinfo {author} {\bibfnamefont {T.~A.~B.}\ \bibnamefont {Kennedy}}, \ and\
  \bibinfo {author} {\bibfnamefont {A.}~\bibnamefont {Kuzmich}},\ }\href@noop
  {} {\bibfield  {journal} {\bibinfo  {journal} {Phys. Rev. A}\ }\textbf
  {\bibinfo {volume} {81}},\ \bibinfo {pages} {041805(R)} (\bibinfo {year}
  {2010})}\BibitemShut {NoStop}%
\bibitem [{\citenamefont {Nunn}\ \emph {et~al.}(2010)\citenamefont {Nunn},
  \citenamefont {Dorner}, \citenamefont {Michelberger}, \citenamefont {Reim},
  \citenamefont {Lee}, \citenamefont {Langford}, \citenamefont {Walmsley},\
  and\ \citenamefont {Jaksch}}]{NDMRLLWJ_2010}%
  \BibitemOpen
  \bibfield  {author} {\bibinfo {author} {\bibfnamefont {J.}~\bibnamefont
  {Nunn}}, \bibinfo {author} {\bibfnamefont {U.}~\bibnamefont {Dorner}},
  \bibinfo {author} {\bibfnamefont {P.}~\bibnamefont {Michelberger}}, \bibinfo
  {author} {\bibfnamefont {K.~F.}\ \bibnamefont {Reim}}, \bibinfo {author}
  {\bibfnamefont {K.~C.}\ \bibnamefont {Lee}}, \bibinfo {author} {\bibfnamefont
  {N.~K.}\ \bibnamefont {Langford}}, \bibinfo {author} {\bibfnamefont {I.~A.}\
  \bibnamefont {Walmsley}}, \ and\ \bibinfo {author} {\bibfnamefont
  {D.}~\bibnamefont {Jaksch}},\ }\href@noop {} {\bibfield  {journal} {\bibinfo
  {journal} {Phys. Rev. A}\ }\textbf {\bibinfo {volume} {82}},\ \bibinfo
  {pages} {022327} (\bibinfo {year} {2010})}\BibitemShut {NoStop}%
\bibitem [{\citenamefont {Jelezko}\ and\ \citenamefont
  {Wrachtrup}(2006)}]{JW_2006}%
  \BibitemOpen
  \bibfield  {author} {\bibinfo {author} {\bibfnamefont {F.}~\bibnamefont
  {Jelezko}}\ and\ \bibinfo {author} {\bibfnamefont {J.}~\bibnamefont
  {Wrachtrup}},\ }\href@noop {} {\bibfield  {journal} {\bibinfo  {journal}
  {Phys. Status Solidi A}\ }\textbf {\bibinfo {volume} {203}},\ \bibinfo
  {pages} {3207} (\bibinfo {year} {2006})}\BibitemShut {NoStop}%
\bibitem [{\citenamefont {Balasubramanian}\ \emph {et~al.}(2009)\citenamefont
  {Balasubramanian}, \citenamefont {Neumann}, \citenamefont {Twitchen},
  \citenamefont {Markham}, \citenamefont {Kolesov}, \citenamefont {Mizuochi},
  \citenamefont {Isoya}, \citenamefont {Achard}, \citenamefont {Beck},
  \citenamefont {Tissler}, \citenamefont {Jacques}, \citenamefont {Hemmer},
  \citenamefont {Jelezko},\ and\ \citenamefont
  {Wrachtrup}}]{BNTMKMIABTJHJW_2009}%
  \BibitemOpen
  \bibfield  {author} {\bibinfo {author} {\bibfnamefont {G.}~\bibnamefont
  {Balasubramanian}}, \bibinfo {author} {\bibfnamefont {P.}~\bibnamefont
  {Neumann}}, \bibinfo {author} {\bibfnamefont {D.}~\bibnamefont {Twitchen}},
  \bibinfo {author} {\bibfnamefont {M.}~\bibnamefont {Markham}}, \bibinfo
  {author} {\bibfnamefont {R.}~\bibnamefont {Kolesov}}, \bibinfo {author}
  {\bibfnamefont {N.}~\bibnamefont {Mizuochi}}, \bibinfo {author}
  {\bibfnamefont {J.}~\bibnamefont {Isoya}}, \bibinfo {author} {\bibfnamefont
  {J.}~\bibnamefont {Achard}}, \bibinfo {author} {\bibfnamefont
  {J.}~\bibnamefont {Beck}}, \bibinfo {author} {\bibfnamefont {J.}~\bibnamefont
  {Tissler}}, \bibinfo {author} {\bibfnamefont {V.}~\bibnamefont {Jacques}},
  \bibinfo {author} {\bibfnamefont {P.~R.}\ \bibnamefont {Hemmer}}, \bibinfo
  {author} {\bibfnamefont {F.}~\bibnamefont {Jelezko}}, \ and\ \bibinfo
  {author} {\bibfnamefont {J.}~\bibnamefont {Wrachtrup}},\ }\href@noop {}
  {\bibfield  {journal} {\bibinfo  {journal} {Nature Materials}\ }\textbf
  {\bibinfo {volume} {8}},\ \bibinfo {pages} {383} (\bibinfo {year}
  {2009})}\BibitemShut {NoStop}%
\bibitem [{\citenamefont {Hemmer}\ \emph {et~al.}(2001)\citenamefont {Hemmer},
  \citenamefont {Turukhin}, \citenamefont {Shahriar},\ and\ \citenamefont
  {Musser}}]{HTSM_2001}%
  \BibitemOpen
  \bibfield  {author} {\bibinfo {author} {\bibfnamefont {P.~R.}\ \bibnamefont
  {Hemmer}}, \bibinfo {author} {\bibfnamefont {A.~V.}\ \bibnamefont
  {Turukhin}}, \bibinfo {author} {\bibfnamefont {M.~S.}\ \bibnamefont
  {Shahriar}}, \ and\ \bibinfo {author} {\bibfnamefont {J.~A.}\ \bibnamefont
  {Musser}},\ }\href@noop {} {\bibfield  {journal} {\bibinfo  {journal} {Optics
  Letters}\ }\textbf {\bibinfo {volume} {26}},\ \bibinfo {pages} {361}
  (\bibinfo {year} {2001})}\BibitemShut {NoStop}%
\bibitem [{\citenamefont {Santori}\ \emph {et~al.}(2006)\citenamefont
  {Santori}, \citenamefont {Tamarat}, \citenamefont {Neumann}, \citenamefont
  {Wrachtrup}, \citenamefont {Fattal}, \citenamefont {Beausoleil},
  \citenamefont {Rabeau}, \citenamefont {Olivero}, \citenamefont {Greentree},
  \citenamefont {Prawer}, \citenamefont {Jelezko},\ and\ \citenamefont
  {Hemmer}}]{STNWFBROGPJH_2006}%
  \BibitemOpen
  \bibfield  {author} {\bibinfo {author} {\bibfnamefont {C.}~\bibnamefont
  {Santori}}, \bibinfo {author} {\bibfnamefont {P.}~\bibnamefont {Tamarat}},
  \bibinfo {author} {\bibfnamefont {P.}~\bibnamefont {Neumann}}, \bibinfo
  {author} {\bibfnamefont {J.}~\bibnamefont {Wrachtrup}}, \bibinfo {author}
  {\bibfnamefont {D.}~\bibnamefont {Fattal}}, \bibinfo {author} {\bibfnamefont
  {R.~G.}\ \bibnamefont {Beausoleil}}, \bibinfo {author} {\bibfnamefont
  {J.}~\bibnamefont {Rabeau}}, \bibinfo {author} {\bibfnamefont
  {P.}~\bibnamefont {Olivero}}, \bibinfo {author} {\bibfnamefont {A.~D.}\
  \bibnamefont {Greentree}}, \bibinfo {author} {\bibfnamefont {S.}~\bibnamefont
  {Prawer}}, \bibinfo {author} {\bibfnamefont {F.}~\bibnamefont {Jelezko}}, \
  and\ \bibinfo {author} {\bibfnamefont {P.}~\bibnamefont {Hemmer}},\
  }\href@noop {} {\bibfield  {journal} {\bibinfo  {journal} {Phys. Rev. Lett.}\
  }\textbf {\bibinfo {volume} {97}},\ \bibinfo {pages} {247401} (\bibinfo
  {year} {2006})}\BibitemShut {NoStop}%
\bibitem [{\citenamefont {de~Lange}\ \emph {et~al.}(2010)\citenamefont
  {de~Lange}, \citenamefont {Wang}, \citenamefont {Rist{\`{e}}}, \citenamefont
  {Dobrovitski},\ and\ \citenamefont {Hanson}}]{LWRDH_2010}%
  \BibitemOpen
  \bibfield  {author} {\bibinfo {author} {\bibfnamefont {G.}~\bibnamefont
  {de~Lange}}, \bibinfo {author} {\bibfnamefont {Z.~H.}\ \bibnamefont {Wang}},
  \bibinfo {author} {\bibfnamefont {D.}~\bibnamefont {Rist{\`{e}}}}, \bibinfo
  {author} {\bibfnamefont {V.~V.}\ \bibnamefont {Dobrovitski}}, \ and\ \bibinfo
  {author} {\bibfnamefont {R.}~\bibnamefont {Hanson}},\ }\href@noop {}
  {\bibfield  {journal} {\bibinfo  {journal} {Science}\ }\textbf {\bibinfo
  {volume} {330}},\ \bibinfo {pages} {60} (\bibinfo {year} {2010})}\BibitemShut
  {NoStop}%
\bibitem [{\citenamefont {Ryan}\ \emph {et~al.}(2010)\citenamefont {Ryan},
  \citenamefont {Hodges},\ and\ \citenamefont {Cory}}]{RHC_2010}%
  \BibitemOpen
  \bibfield  {author} {\bibinfo {author} {\bibfnamefont {C.~A.}\ \bibnamefont
  {Ryan}}, \bibinfo {author} {\bibfnamefont {J.~S.}\ \bibnamefont {Hodges}}, \
  and\ \bibinfo {author} {\bibfnamefont {D.~G.}\ \bibnamefont {Cory}},\
  }\href@noop {} {\bibfield  {journal} {\bibinfo  {journal} {Phys. Rev. Lett.}\
  }\textbf {\bibinfo {volume} {105}},\ \bibinfo {pages} {200402} (\bibinfo
  {year} {2010})}\BibitemShut {NoStop}%
\bibitem [{\citenamefont {Naydenov}\ \emph {et~al.}(2011)\citenamefont
  {Naydenov}, \citenamefont {Dolde}, \citenamefont {Hall}, \citenamefont
  {Shin}, \citenamefont {Fedder}, \citenamefont {Hollenberg}, \citenamefont
  {Jelezko},\ and\ \citenamefont {Wrachtrup}}]{NDHSFHJW_2011}%
  \BibitemOpen
  \bibfield  {author} {\bibinfo {author} {\bibfnamefont {B.}~\bibnamefont
  {Naydenov}}, \bibinfo {author} {\bibfnamefont {F.}~\bibnamefont {Dolde}},
  \bibinfo {author} {\bibfnamefont {L.~T.}\ \bibnamefont {Hall}}, \bibinfo
  {author} {\bibfnamefont {C.}~\bibnamefont {Shin}}, \bibinfo {author}
  {\bibfnamefont {H.}~\bibnamefont {Fedder}}, \bibinfo {author} {\bibfnamefont
  {L.~C.~L.}\ \bibnamefont {Hollenberg}}, \bibinfo {author} {\bibfnamefont
  {F.}~\bibnamefont {Jelezko}}, \ and\ \bibinfo {author} {\bibfnamefont
  {J.}~\bibnamefont {Wrachtrup}},\ }\href@noop {} {\bibfield  {journal}
  {\bibinfo  {journal} {Phys. Rev. B}\ }\textbf {\bibinfo {volume} {83}},\
  \bibinfo {pages} {081201} (\bibinfo {year} {2011})}\BibitemShut {NoStop}%
\bibitem [{\citenamefont {Dutt}\ \emph {et~al.}(2007)\citenamefont {Dutt},
  \citenamefont {Childress}, \citenamefont {Jiang}, \citenamefont {Togan},
  \citenamefont {Maze}, \citenamefont {Jelezko}, \citenamefont {Zibrov},
  \citenamefont {Hemmer},\ and\ \citenamefont {Lukin}}]{GCJTMJZHL_2007}%
  \BibitemOpen
  \bibfield  {author} {\bibinfo {author} {\bibfnamefont {M.~V.~G.}\
  \bibnamefont {Dutt}}, \bibinfo {author} {\bibfnamefont {L.}~\bibnamefont
  {Childress}}, \bibinfo {author} {\bibfnamefont {L.}~\bibnamefont {Jiang}},
  \bibinfo {author} {\bibfnamefont {E.}~\bibnamefont {Togan}}, \bibinfo
  {author} {\bibfnamefont {J.}~\bibnamefont {Maze}}, \bibinfo {author}
  {\bibfnamefont {F.}~\bibnamefont {Jelezko}}, \bibinfo {author} {\bibfnamefont
  {A.~S.}\ \bibnamefont {Zibrov}}, \bibinfo {author} {\bibfnamefont {P.~R.}\
  \bibnamefont {Hemmer}}, \ and\ \bibinfo {author} {\bibfnamefont {M.~D.}\
  \bibnamefont {Lukin}},\ }\href@noop {} {\bibfield  {journal} {\bibinfo
  {journal} {Science}\ }\textbf {\bibinfo {volume} {316}},\ \bibinfo {pages}
  {1312} (\bibinfo {year} {2007})}\BibitemShut {NoStop}%
\bibitem [{\citenamefont {Maurer}\ \emph {et~al.}(2012)\citenamefont {Maurer},
  \citenamefont {Kucsko}, \citenamefont {Latta}, \citenamefont {Jiang},
  \citenamefont {Yao}, \citenamefont {Bennett}, \citenamefont {Pastawski},
  \citenamefont {Hunger}, \citenamefont {Chisholm}, \citenamefont {Markham},
  \citenamefont {Twitchen}, \citenamefont {Cirac},\ and\ \citenamefont
  {Lukin}}]{MKLJYBPHCMTCL_2012}%
  \BibitemOpen
  \bibfield  {author} {\bibinfo {author} {\bibfnamefont {P.~C.}\ \bibnamefont
  {Maurer}}, \bibinfo {author} {\bibfnamefont {G.}~\bibnamefont {Kucsko}},
  \bibinfo {author} {\bibfnamefont {C.}~\bibnamefont {Latta}}, \bibinfo
  {author} {\bibfnamefont {L.}~\bibnamefont {Jiang}}, \bibinfo {author}
  {\bibfnamefont {N.~Y.}\ \bibnamefont {Yao}}, \bibinfo {author} {\bibfnamefont
  {S.~D.}\ \bibnamefont {Bennett}}, \bibinfo {author} {\bibfnamefont
  {F.}~\bibnamefont {Pastawski}}, \bibinfo {author} {\bibfnamefont
  {D.}~\bibnamefont {Hunger}}, \bibinfo {author} {\bibfnamefont
  {N.}~\bibnamefont {Chisholm}}, \bibinfo {author} {\bibfnamefont
  {M.}~\bibnamefont {Markham}}, \bibinfo {author} {\bibfnamefont {D.~J.}\
  \bibnamefont {Twitchen}}, \bibinfo {author} {\bibfnamefont {J.~I.}\
  \bibnamefont {Cirac}}, \ and\ \bibinfo {author} {\bibfnamefont {M.~D.}\
  \bibnamefont {Lukin}},\ }\href@noop {} {\bibfield  {journal} {\bibinfo
  {journal} {Science}\ }\textbf {\bibinfo {volume} {336}},\ \bibinfo {pages}
  {1283} (\bibinfo {year} {2012})}\BibitemShut {NoStop}%
\end{thebibliography}%

\end{document}